\documentstyle[11pt,times,emulateapj,psfig]{article} %
\submitted{Accepted by the Astrophysical Journal, Part I}
\newcommand{\sna}{SNe~Ia}
\newcommand{\kms}{{\,\rm km}\ {\rm~s}^{-1}} 
\newcommand{\ksm}{{\,\rm km}\ {\rm~s}^{-1}\ {\rm~Mpc}^{-1}} 

\lefthead{SN~1989B in NGC~3627}   
\righthead{Saha et al.}
  
\begin{document}
\title{Cepheid Calibration of the Peak Brightness of \sna. \\ IX.
       SN~1989B in NGC~3627\footnotemark}
\footnotetext{Based on observations with the NASA/ESA {\it Hubble Space 
Telescope}, obtained at the Space Telescope Science Institute, operated 
by AURA, Inc. under NASA contract NAS5-26555.}

\author{A. Saha} 
\affil{National Optical Astronomy Observatories, 
950 North Cherry Ave., Tucson, AZ 85726} 
\author{Allan Sandage}
\affil{Observatories of the Carnegie Institution of Washington, 
813 Santa Barbara Street, Pasadena, CA 91101} 
\author{ G.A. Tammann and Lukas Labhardt} 
\affil{Astronomisches Institut der Universit\"at Basel,  
Venusstrasse 7, CH-4102 Binningen, Switzerland} 
\and
\author{F.D. Macchetto\altaffilmark{2} and N. Panagia\altaffilmark{2}}
\affil{Space Telescope Science Institute, 3700 San Martin Drive,
Baltimore, MD 21218}
\altaffiltext{2}{Affiliated to the Astrophysics Division, Space 
Sciences Department of ESA.}

\begin{abstract}
Repeated imaging observations have been made of NGC~3627 with the {\it
Hubble Space Telescope} in 1997/98, over
an interval of 58 days.  Images were obtained on 12 epochs in the
$F555W$ band and on five epochs in the $F814W$ band.  The galaxy
hosted the prototypical, ``Branch normal'', type Ia supernova
SN~1989B.  A total of 83 variables have been found, of which 68 are
definite Cepheid variables with periods ranging from 75 days to 3.85
days.  The de-reddened distance modulus is determined to be $(m -M)_0 =
30.22 \pm 0.12$ (internal uncertainty) using a subset of
the Cepheid data whose reddening and error parameters are secure.

The photometric data of \cite{wel94} (1994), combined with the Cepheid data
for NGC~3627 give $M_B({\rm max}) = -19.36 \pm 0.18$ and $M_V({\rm
max}) = -19.34 \pm 0.16$ for SN~1989B.  Combined with the previous six
calibrations in this program, plus two additional calibrations
determined by others gives the mean absolute magnitudes at maximum of
$\langle{M_B}\rangle = -19.48 \pm 0.07$ and $\langle{M_V}\rangle =
-19.48 \pm 0.07$ (Table~\ref{tbl5}) for ``Branch normal'' SNe~Ia at
this interim stage in the calibration program. 

Using the argument by \cite{wel94} (1994) that SN~1989B here is virtually
identical in decay rate and colors at maximum with SN~1980N in NGC~1316
in the Fornax cluster, and that such identity means nearly identical
absolute magnitude, it follows that the difference in the
distance modulus of NGC~3627 and NGC~1316 is $1.62 \pm 0.03$ mag.  
Thus the NGC~3627 modulus implies that 
$(m - M)_0 = 31.84$ for NGC~1316. 

The second parameter correlations of $M$(max) of blue SNe~Ia with
decay rate, color at maximum, and Hubble type are re-investigated.
The dependence of $\langle{M{\rm(max)}}\rangle$ on decay rate is
non-linear, showing a minimum for decay rates between $1.0 < \Delta
m_{\rm 15} < 1.6$.  Magnitudes corrected for decay rate show no
dependence on Hubble type, but a dependence on color remains.
Correcting both the fiducial sample of 34 SNe~Ia with decay-rate data
and the current eight calibrating SNe~Ia for the correlation with
decay rate as well as color gives
$H_0 = 60 \pm 2~{\rm (internal)}\ksm$,
in both $B$ and $V$.  The same value to within 4\% is obtained if only
the SNe~Ia in spirals (without second parameter corrections) are
considered.

The correlation of SNe~Ia color at maximum with $M$(max) cannot be due
to internal absorption because the slope coefficients in $B$, $V$, and
$I$ with the change in magnitude are far from or even opposite to the
canonical reddening values.  The color effect must be intrinsic to the
supernova physics.  ``Absorption'' corrections of distant blue SNe~Ia
will lead to incorrect values of $H_0$.

The Cepheid distances used in this series are insensitive to
metallicity differences (\cite{sanetal99} 1999). The zeropoint of the P-L
relation is based on an assumed LMC modulus of $(m - M)_0 = 18.50$.
As this may have to be increased by $0\fm06$ to $0\fm08$, all distances
in this paper will follow and $H_0$ will decrease by 3 -- 4\%.
\end{abstract}


\section{Introduction} 
 
This is the ninth paper of a series whose purpose is to obtain Cepheid
distances to galaxies that have produced supernovae of type Ia
(SNe~Ia), thereby calibrating their absolute magnitudes at maximum
light.  The Hubble diagram for SNe~Ia that are not abnormal 
in either their intrinsic colors or their spectra at maximum
(\cite{bra93} 1993) is exceedingly tight (\cite{sata93} 1993; 
\cite{tasa95} 1995; \cite{saha97} 1997 ; \cite{par99} 1999) 
even before second-order corrections for light curve decay rate 
or color (\cite{ham95} 1995, 1996a, b; \cite{rie96} 1996)
are applied.  Hence, the absolute magnitude calibrations lead directly
to a good estimate
of the global value of the Hubble constant because the
SNe~Ia Hubble diagram is defined at redshifts that are well beyond any
local velocity anomalies in the Hubble flow.

The previous papers of this series concern Cepheids in IC~4182 for
SN~1937C (\cite{sanetal92} 1992, Paper~I; \cite{saha94} 1994,
Paper~II), NGC~5253 for the two SNe~Ia 1895B and 1972E
(\cite{sanetal94} 1994, Paper~III; \cite{saha95} 1995, Paper~IV),
NGC~4536 for SN~1981B (\cite{saha96a} 1996a, Paper~V); NGC~4496A for
SN~1960F (\cite{saha96b} 1996b, Paper~VI); NGC~4639 for SN~1990N 
(\cite{sanetal96} 1996, Paper~VII; \cite{saha97} 1997, Paper~VIII).

The purpose of this paper is to set out the data for the discovery and
photometry of Cepheids in NGC~3627, the parent galaxy of the type Ia
SN~1989B. This case, together with the new SN~Ia 1998bu in NGC~3368
for which \cite{tan95} (1995) have a Cepheid distance, and the new Cepheid
distance of NGC~4414 (\cite{tur98} 1998), parent galaxy to SN~1974G, and
the compilation of the extant photometric data on SN~1998bu
(\cite{sun98} 1998) and on SN~1974G (\cite{schae98} 1998), now increase the
number of SNe~Ia calibrators from seven in \cite{sanetal96} (1996) and
\cite{saha97} (1997) to nine here. It will be recalled that we have used the
absolute magnitude of SN~1989B in two previous discussions
(\cite{sanetal96} 1996; \cite{saha97} 1997) but based on the then unproven
premise that the \cite{tan95} (1995) Cepheid distance to NGC~3368 would be
the same as the distance to NGC~3627, as both galaxies are in the
loose Leo association.

SN~1989B was discovered by visual inspection of NGC~3627 by
\cite{eva89} (1989) on January 30, 1989 which was 7 days before maximum light
in $B$.  A detailed light curve in {\it UBVRI} and a determination of
the reddening of $E(B-V) = 0\fm37 \pm 0\fm03$ was made by \cite{wel94} (1994),
following the earlier analysis by \cite{bar90} (1990).  The magnitudes at
maximum corrected for the extinction are $B({\rm
max}) = 10\fm86 \pm 0\fm13$ and $V({\rm max}) = 10\fm88 \pm 0\fm10$.

The parent galaxy NGC~3627 (M~66) is one of the brightest spirals
(Sb(s)II.2) in the complicated region of the Leo group, first isolated
by \cite{hum56} (1956, Table~XI) where 18 possible members were
identified including NGC~3627 and NGC~3368 (M~96). Both galaxies are
illustrated in the Hubble Atlas (\cite{san61} 1961, panels 12 and 23), and
the Carnegie Atlas (\cite{sanbed94} 1994, Panels 118, 137, and S14).

\cite{dev75} (1975) has divided the larger Leo group complex into three
groups.  G9 in his Table~3 is a spiral--dominated subgroup near
NGC~3627 (M~66). His G11 is dominated by NGC~3368 (M~96), parent to
SN~1998bu
and contains NGC~3351, for which a Cepheid distance is also
available (\cite{gra97} 1997).  The agreement of the Cepheid distances
of NGC~3368 and NGC~3351 to within $0\fm36 \pm 0\fm25$ is
satisfactory.  The smaller group G49 surrounds NGC~3607.  The division
into three subgroups is supported by the three-dimensional
hierarchical clustering analysis of \cite{mat78} (1978).  A catalog of 52
possible members of the subgrouping in the field of NGC~3368,
NGC~3379, and NGC~3384 is given by \cite{ferg90} (1990).

Because of the complication of three subgroups rather than a well
defined single group, we have never been totally convinced that the
distance to NGC~3627, required to calibrate SN~1989B, could be taken
to be that of the Cepheids in NGC~3368 from \cite{tan95} (1995), although we
analyzed our calibration data in Papers~VII and VIII on that
premise. This assumption is now no longer necessary with the discovery
and analysis of Cepheids in NGC~3627.

\begin{figure*}
\centerline{\hbox{\psfig{figure=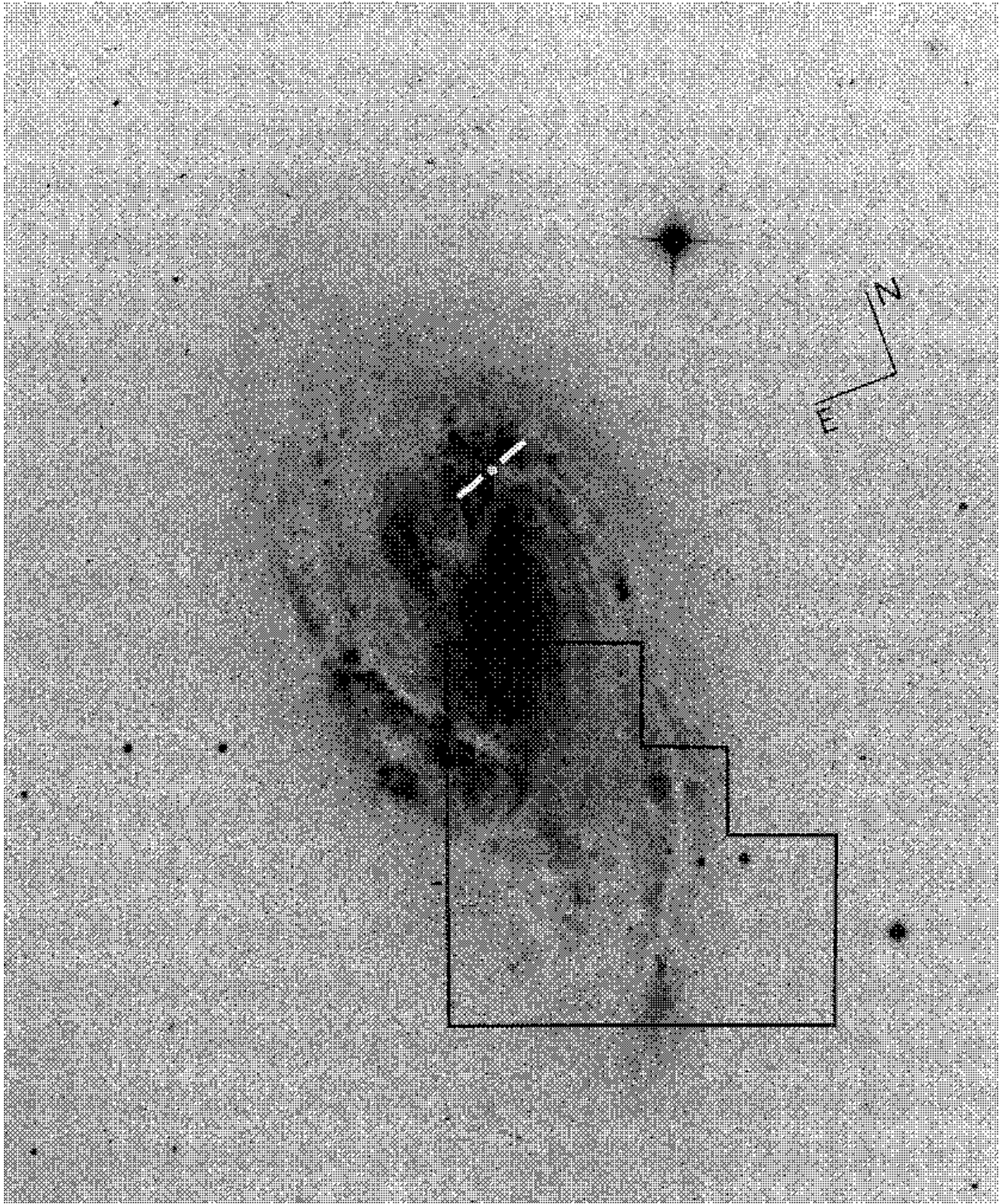,width=17.0cm}}}
\caption{Ground-based image of NGC~3627 made with the Mount
Wilson Hooker 100-inch reflector by Hubble on November 29/30,
1946. The image is from the same original negative as used for the
short exposure image as an insert in The Carnegie Atlas, Panel~137.
The field covered by the WFPC2 of the $HST$ is superposed.  The
position of SN~1989B (\cite{bar90} 1990) is marked.\label{fig1}}
\end{figure*}

Figure~\ref{fig1} shows a ground-based image of NGC~3627 made from the
Mount Wilson 100-inch Hooker blue plate (E40) that was used in the
Hubble Atlas and for one of the frames of the Carnegie Atlas.  The
position of the four WFPC2 chips of the $HST$ is superposed to show
our search area for the Cepheids. The position of SN~1989B is marked,
taken from the photographs in \cite{bar90} (1990).

\begin{figure*}
\centerline{\hbox{\psfig{figure=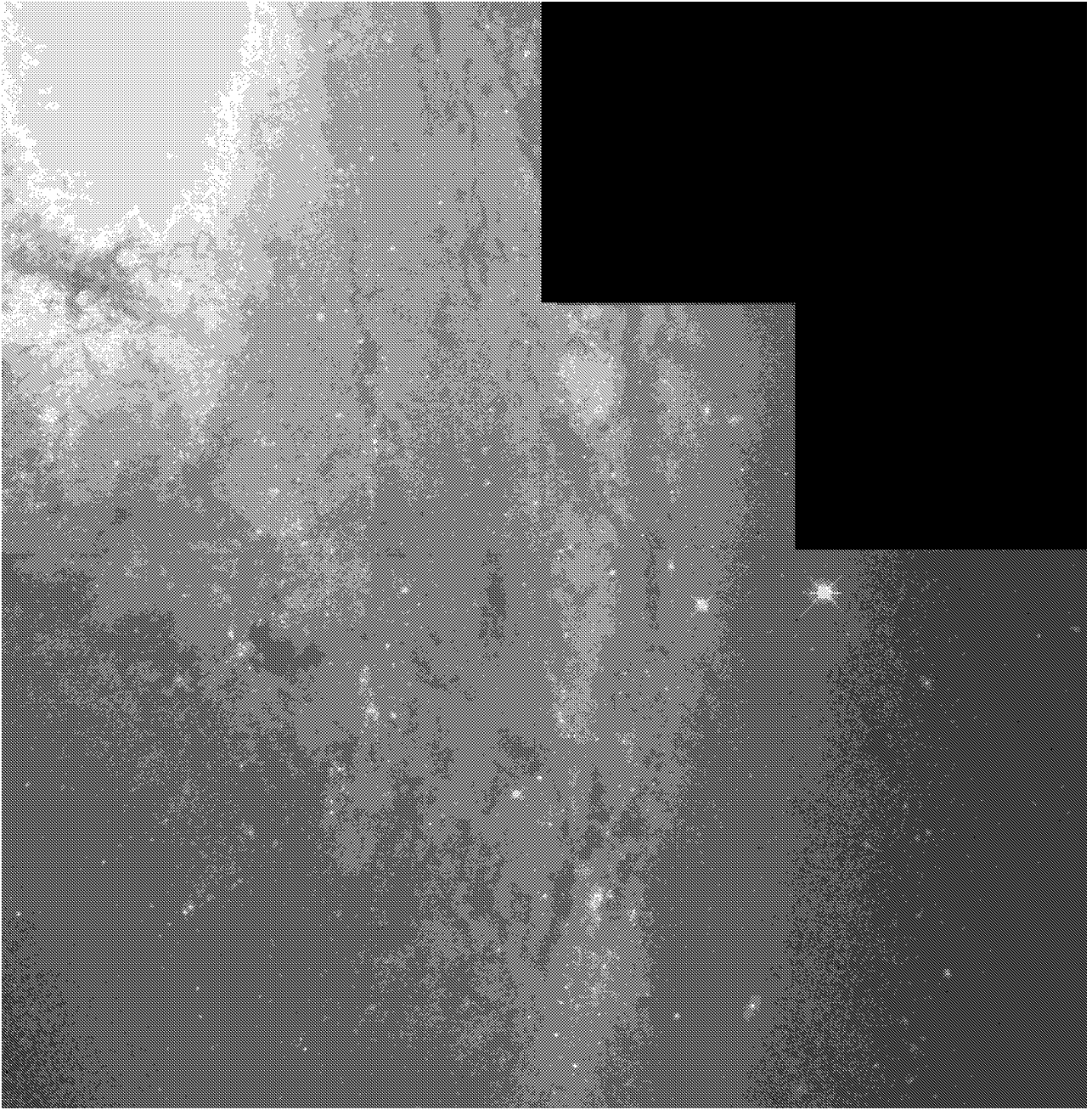,width=17.0cm,angle=180 }}}
\caption{Color image of the $HST$ field made by stacking several
frames in both the $V$ and $I$ passbands and combining.
\label{fig2}}
\end{figure*}

Figure~\ref{fig2} is a color composite $HST$ montage of stacked
frames.  The dust lanes are striking both here and in Fig.~\ref{fig1}.
The extinction clearly varies over short scales on the image. This
forewarns about the importance of differential extinction to the
Cepheids, requiring good color information for them to obtain a
reliable true modulus. The extinction in NGC~3627 is more severe than
in any galaxy of our six previous calibrations in Papers~I--VIII of
the series.

The journal of the $HST$ observations and the photometry of the master
template frame are in the next section. The identification and
classification of the variables are in \S~3. The apparent
period-luminosity relations, the analysis of the severe absorption
problem and how we have corrected for it, the resulting distance
modulus, and the absolute magnitude at maximum for SN~1989B are in
\S~4. In \S~5 we combine the data with our six previous calibrators,
as well as SN~1974G in NGC~4414 and SN~1998bu in NGC~3368 from
external sources.  
This paper provides a direct Cepheid distance to NGC~3627 and removes the 
uncertainty, thereby strengthening the weight of SN1989B in the overall 
calibration.
A first value of $H_0$ is given.  A discussion of
the second-parameter corrections and their effect on $H_0$ is in the
penultimate section \S~6.

\section{Observations and Photometry} 
\subsection{The Data} 

Repeated images of the field in NGC~3627, shown in Fig.~\ref{fig2}, were
obtained using the WFPC2 (\cite{hol95a} 1995a) on the $HST$
between November 1997 and January 1998.  There are 12
discrete epochs in the $F555W$ passband, and 5 epochs in the $F814W$
passband, spanning a period of 58 days. The duration of this period is
constrained by the time window during which this target can be
observed with $HST$ without altering the field orientation. This
window was further curtailed due to logistics for accommodating a
campaign with the NICMOS camera, for which a change of the telescope
focus was necessary, thus rendering the telescope useless for other
kinds of observations.

The epochs were spaced strategically over this period to provide
maximum leverage on detecting and finding periods of Cepheid variables
over the period range 10 to 60 days. Each epoch in each filter was
made of two sub-exposures taken back-to-back on successive orbits of
the spacecraft. This allows the removal of cosmic rays by an
anti-coincidence technique. The images from various epochs were
co-aligned to within 3--4 pixels on the scale of the PC chip, which is
1--2 pixels on the scale of the other three wide-field chips. The
journal of observations is given in Table~\ref{tbl1}.

\subsection{Photometry} 

The details of processing the images, combining the sub-exposures for
each epoch while removing cosmic rays and performing the photometry
with a variant of DoPHOT (\cite{schec93} 1993) optimized for WFPC2 data has
been given in \cite{saha96a} (1996a) and need not be repeated here. The data
reduction procedure is identical to that described in Paper~V, with
the one exception of a change in the definition of the ``partial
aperture''. Instead of the $9 \times 9$ pixel aperture, a circular
aperture of 5 pixel radius was used, and the local background is
defined as the value for which the aperture growth curve is flat from
6 to 8 pixels. The details of these changes are given in \cite{ste98} (1998)
in their \S~2.2. This change produces no known systematic differences, but 
improves the S/N with which aperture corrections are measured.

In keeping with the precepts in Paper~V, the measurements in any one
passband are expressed in the magnitude system defined by
\cite{hol95b} (1995b) that is native to the WFPC2.  These are the $F555W$ and
$F814W$ ``ground system'' magnitudes calibrated with $HST$ ``short''
exposure frames.  The issue of the discrepancy of the photometric
zero-points for the ``long'' and ``short'' WFPC2 exposures, originally
found by \cite{ste95} (1995) is described in some detail in Paper~V.  In any
eventual accounting for this zero-point correction, one must {\it add}
$0\fm05$ in {\it both} passbands to the \cite{hol95b} (1995b) calibration
whenever the exposures are longer than several hundred seconds. The
cause of this zero-point difference is not fully understood at the time
of this writing, and, as in previous papers of this series, we
continue to present the basic photometry (Tables~\ref{tbl3} and
\ref{tbl4}) on the ``uncorrected'' \cite{hol95b} (1995b) ``short exposure''
calibration.  And, because all of our WFPC2 observations have exposure
times that are ``long'', we make the $0\fm05$ adjustment only at the
resulting distance modulus of NGC~3627 (cf. \S~4.2.2.).
Correspondingly the distance moduli in Table~\ref{tbl5} are corrected
to the `long' calibration scale
by $0\fm05$, except the moduli of IC~4182 and NGC~5253 which were
observed with the older WF/PC.

\section{Identification and Classification of the Variable Stars} 

Armed with measured magnitudes and their reported errors at all
available epochs for each star in the object list, the method
described by \cite{saha90} (1990) was used to identify variable stars.  
The details specific to WFPC2 data have been given in various degrees 
of detail in Papers~V, VI, and VIII.

\begin{figure*}
\centerline{\hbox{\psfig{figure=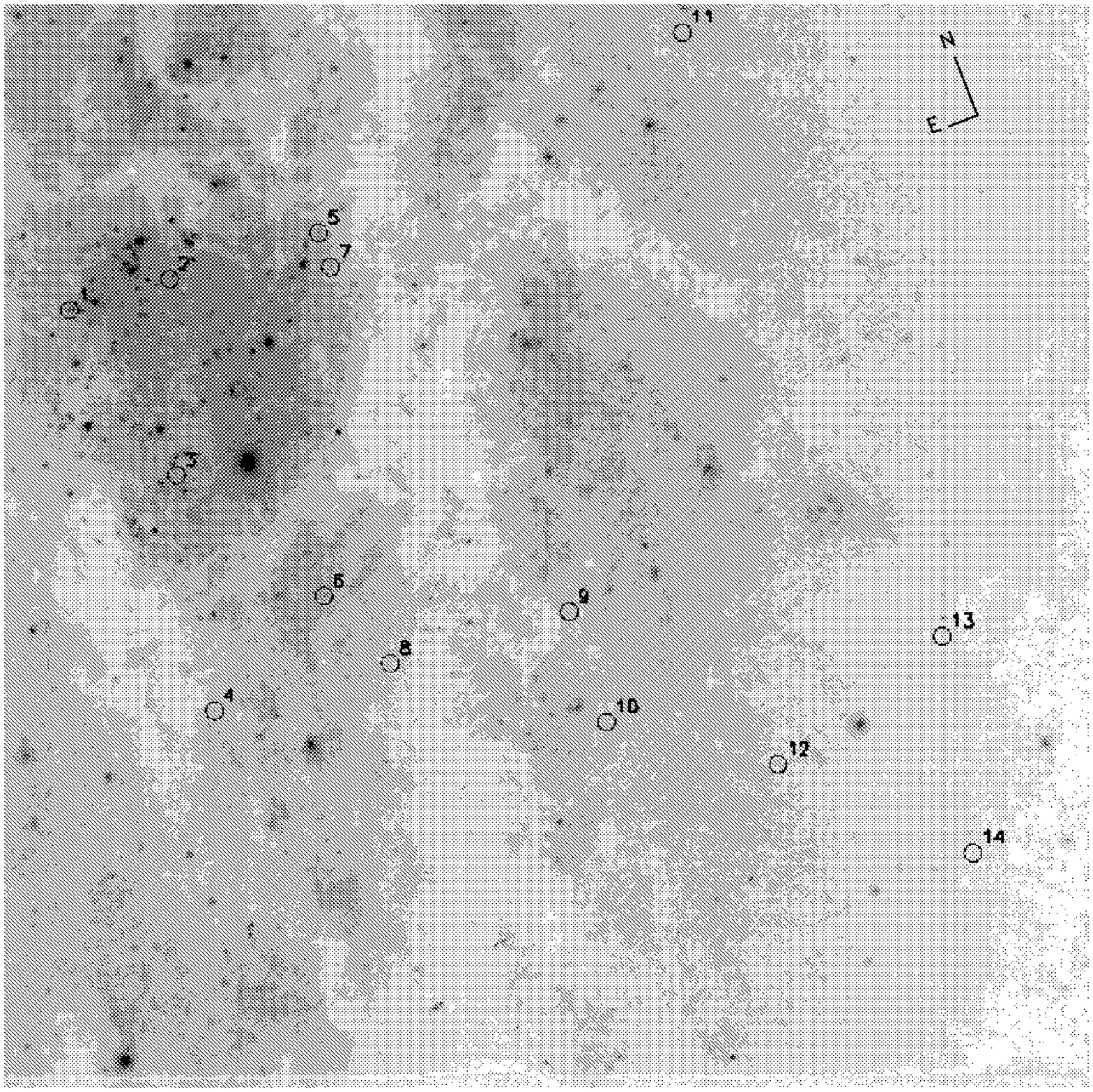,height=8.5cm}
\psfig{figure=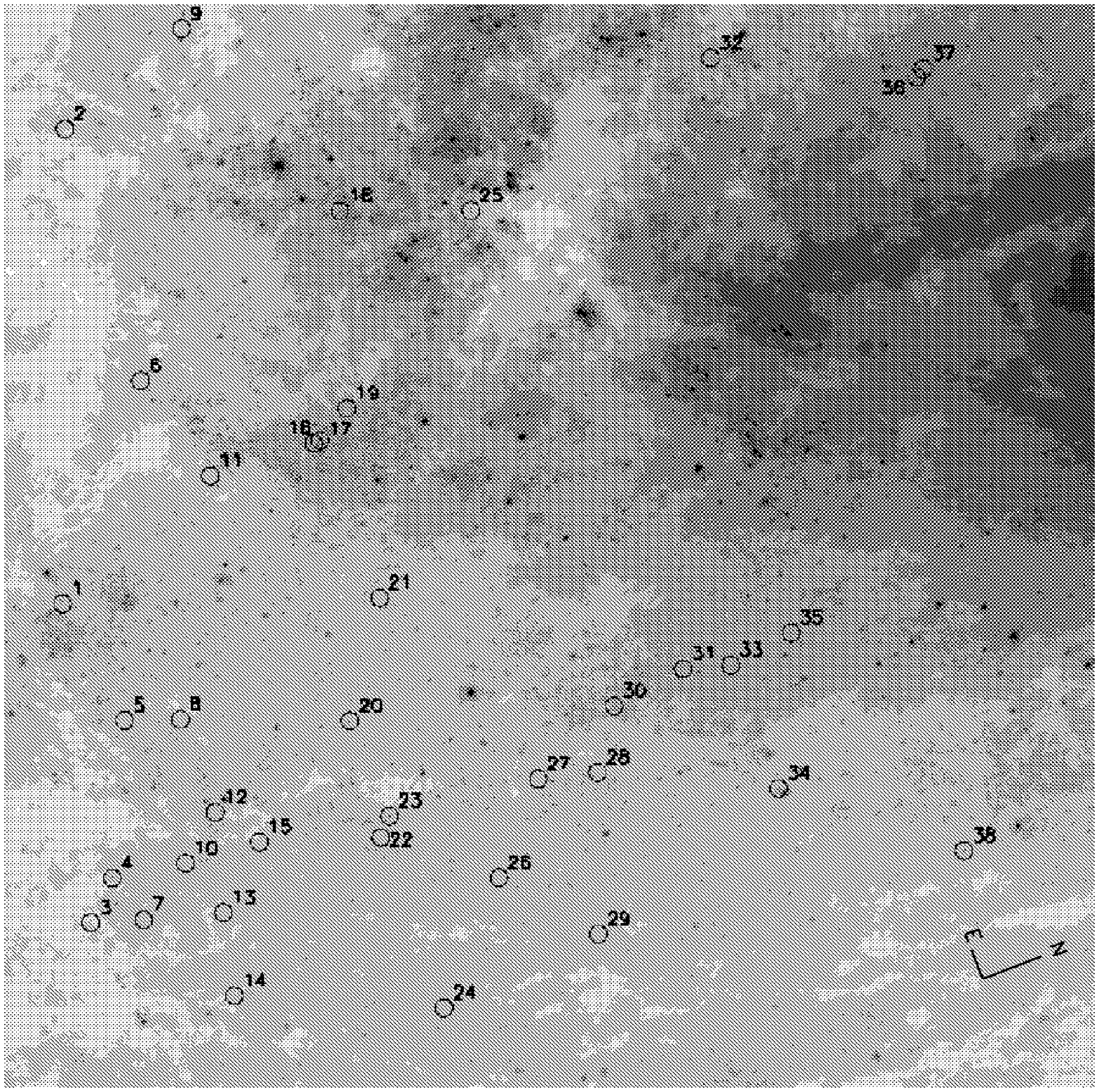,height=8.5cm}}}
\centerline{\hbox{\psfig{figure=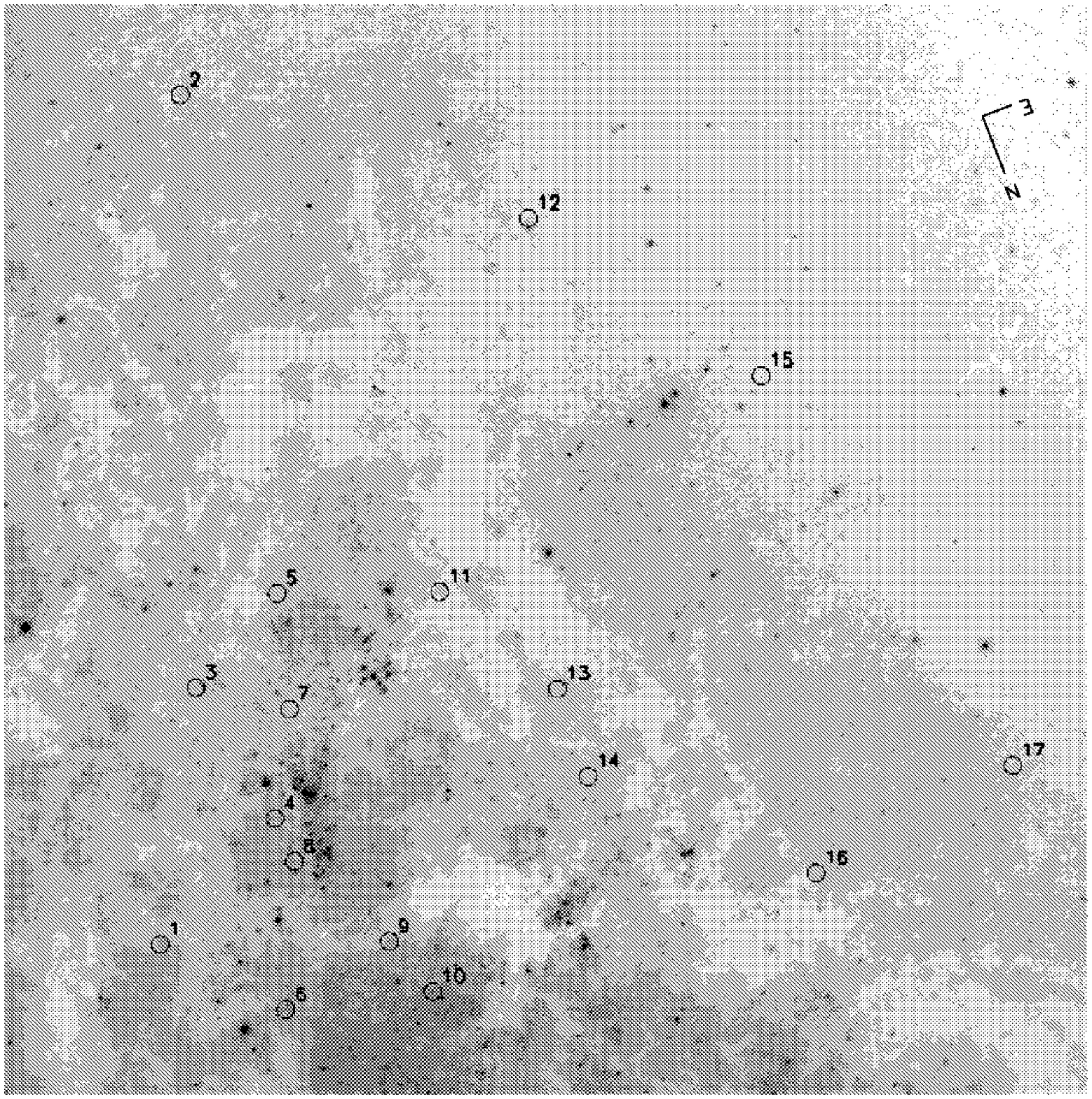,height=8.5cm}
\psfig{figure=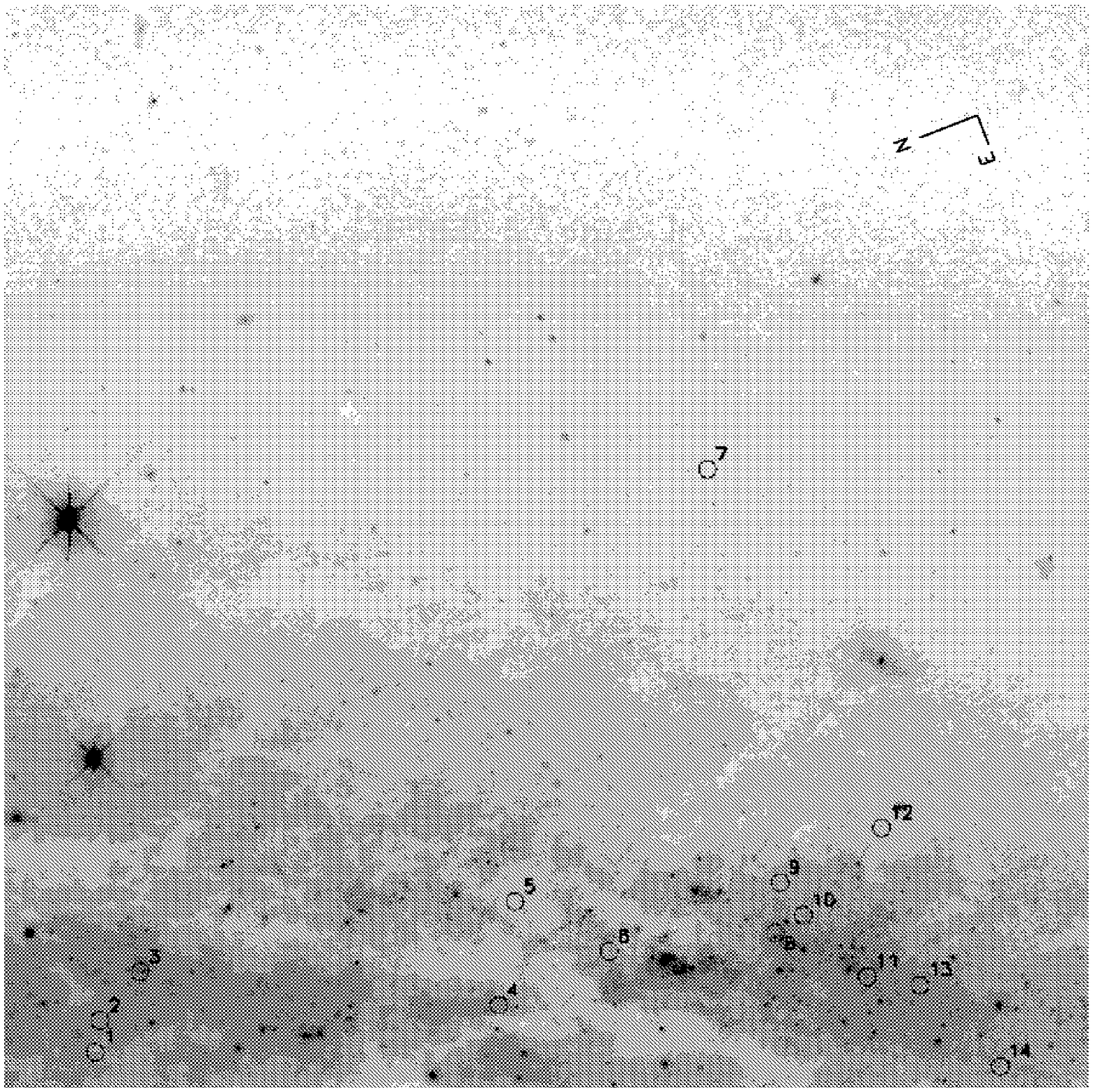,height=8.5cm}}}
\caption{Identifications for all the variable stars found.  The
numbers are the same as in Tables~\ref{tbl3} and \ref{tbl4}.  Each of
the four WFPC2 chips are shown separately.
\label{fig3}}
\end{figure*}

All variable stars definitely identified are marked in
Fig.~\ref{fig3}.  However, some of the identified variables cannot be
seen in Fig.~\ref{fig3} because of their extreme faintness and/or
because of the large variation in surface brightness over the field.
Hence, to complement these charts, we set out in Table~\ref{tbl2} the
X and Y pixel positions for all variable stars as they appear in the
images identified in the $HST$ data archive as U3510701R and
U3510702R.

The photometry on the \cite{hol95b} (1995b) ``short exposure''
calibration system for the final list of 83 variable stars is
presented in Table~\ref{tbl3} for each epoch and each filter.  The
periods were determined with the \cite{laf65} (1965) by using only the
$F555W$ passband data.  Aliassing is not a serious problem for periods
between 10 and 58 days because the observing strategy incorporated an
optimum timing scheme as before in this series.

\begin{figure*}
\centerline{\hbox{\psfig{figure=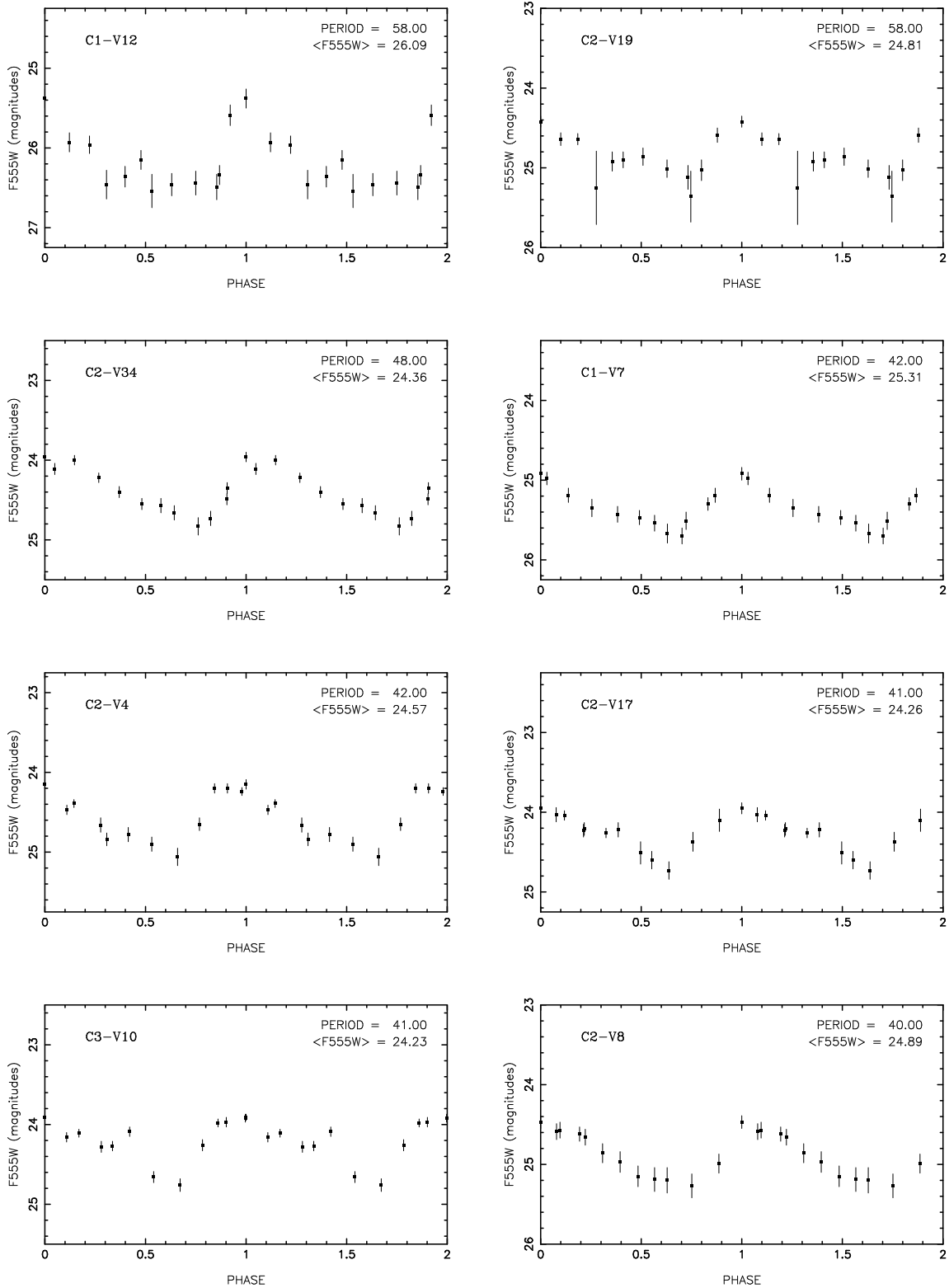,width=16.0cm}}}
\caption{Light curves, plotted in order of period, in the $F555W$
band.  The first four variables are plotted with the arbitrary period
of 100 days to facilitate the visualization.  The actual periods are
unknown; the time interval of the observations of 58 days is shorter
than the actual periods, if any, of these four stars. \label{fig4}}
\end{figure*}

The resulting light curves in the $F555W$ passband, together with
periods and mean magnitudes (determined by integrating the light
curves, converted to intensities, and then converting the average back
to magnitudes, and called the ``phase-weighted intensity average'' in
\cite{saha90} 1990), are shown in Fig.~\ref{fig4}, plotted in the
order of descending period.

Four objects, C2-V23, C2-V25, C2-V36, and C2-V37 are definitely
variable but are unlikely to be Cepheids.  They may be periodic with
periods greater than the time spanned by our 12 epochs, or they may be
transient variables such as novae.  They are plotted in
Fig.~\ref{fig4} with periods set artificially at 100 days (much larger
than the observing time base of 58 days) for the purpose of
visualization.  Four obvious objects, C1-V1, C2-V38, C4-V3, and C4-V14
have periods that are just a little larger than the observing time
base.  Best guesses of the periods have been made from the light curve
shapes.  As the periods of these variables are not definitive, they
should be used with caution in deriving distances.  The remaining
variables have light curves and periods that are consistent with being
Cepheids.

\begin{figure*}
\centerline{\hbox{\psfig{figure=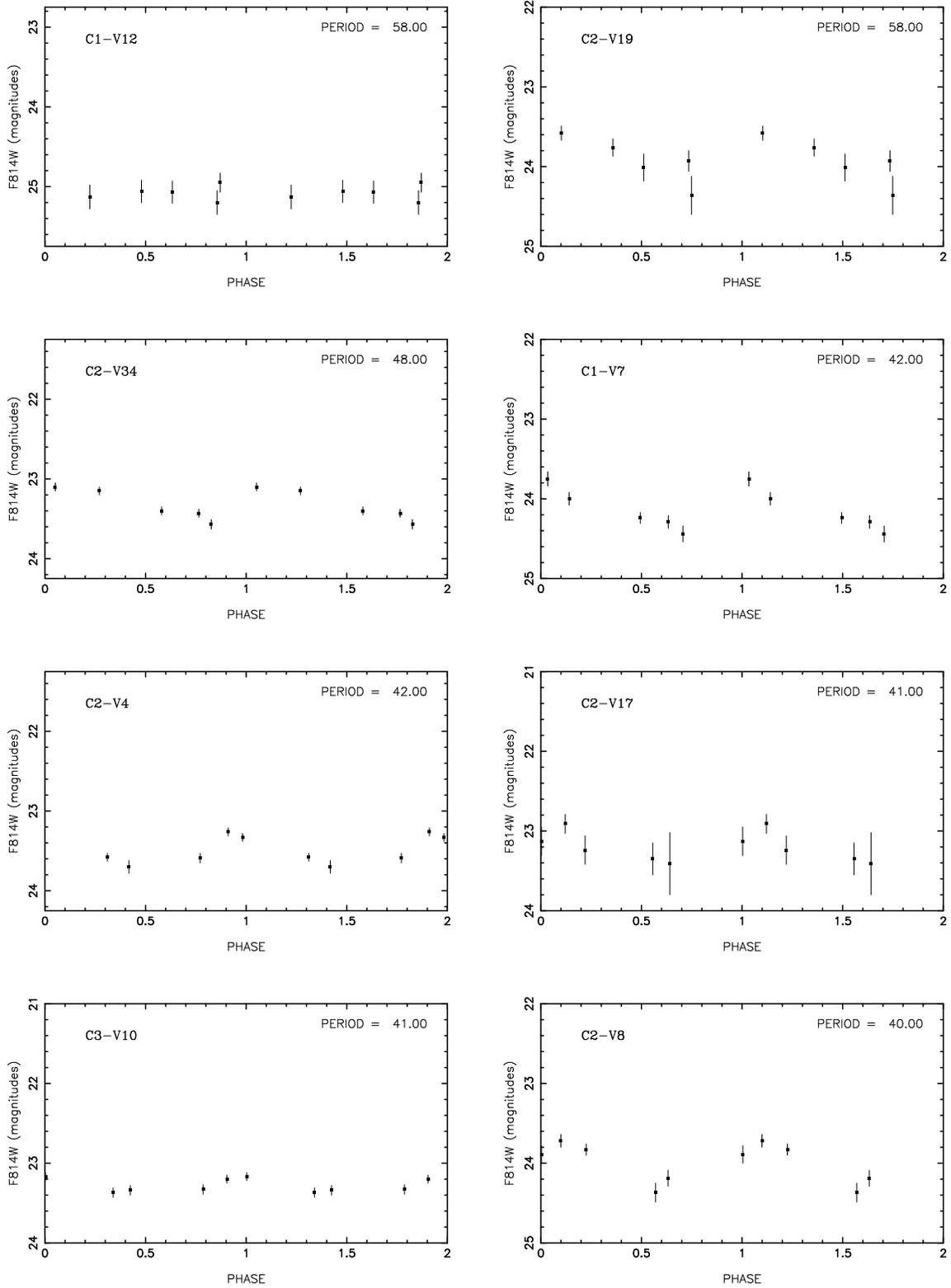,width=16.0cm}}}
\caption{Same as Fig.~\ref{fig4} but for the $F814W$ passband,
adopting the periods and the phasing used in Fig.~\ref{fig4}.
\label{fig5}}
\end{figure*}

The available data for the variables in $F814W$ were folded with the
ephemerides derived above using the $F555W$ data.  The results are
plotted in Fig.~\ref{fig5}.  The long-period variables, transient
variables, and long-period Cepheids (as discussed above) are shown
with the same assigned ephemerides as in Fig.~\ref{fig4}.  Not all the
variables discovered from the $F555W$ photometry were found in the
$F814W$ images. The fainter variables, either because they are
intrinsically faint or else appear faint due to high extinction, may
not register clearly on the $F814W$ frames which do not reach as faint
a limiting magnitude as those in $F555W$.  Since photometry of such
objects was obviously impossible in $F814W$, these variables are
dropped from Fig.~\ref{fig5} and also from further analysis.  Only the
68 variables from the $F555W$ frames that were recovered in at least
one of the $F814W$ epochs are considered further.

The mean magnitudes in $F814W$ (integrated as intensities over the
cycle) were obtained from the procedure of \cite{lab97} (1997) whereby
each $F814W$ magnitude at a randomly sampled phase is converted to a
mean value $\langle{F814W}\rangle$ using amplitude and phase
information from the more complete $F555W$ light curves.  Note that
each available observation of $F814W$ can be used independently to
derive a mean magnitude.  Hence, the scatter of the individual values
about the adopted mean $F814W$ value is an {\it external} measure of
the uncertainty in determining $\langle{F814W}\rangle$.  This value is
retained as the error in $\langle{F814W}\rangle$, and propagated in
the later calculations.

The prescription given in Paper~V for assigning the light-curve
quality index QI (that ranges from 0 to 6) was used.  In this scheme,
two points are given for the quality of the $F555W$ light curves, two
points for the evenness in phase coverage of the five $F814W$
observation epochs, and three points for the amplitude and phase
coherence of the $F814W$ observations compared with the $F555W$ light
curve.  Hence, a quality index of 6 indicates the best possible light
curve quality. A quality index of 2 or less indicates near fatal flaws
such as apparent phase incoherence in the two passbands.  This is
generally the indication that object confusion by crowding and/or
contamination by background is likely.

Table~\ref{tbl4} lists the characteristics of the 68 objects whose
light curves in $F555W$ are consistent with those of Cepheids, and for
which an $F814W$ measurement exists for at least one epoch.  The
$F555W$ and $F814W$ instrumental magnitudes of Table~\ref{tbl3} have
been converted to the Johnson $V$ and Cousins (Cape) $I$ standard
photometric system by the color equations used in Papers V to VIII of
this series, as set out in equations (2) and (3) of Paper~V, based on
the transformations of \cite{hol95b} (1995b).

The magnitude scatter $\sigma_{\langle{V}\rangle}$ in Table~\ref{tbl4}
is based on the estimated measuring errors in the photometry of the
individual epochs.  The determination of the scatter
$\sigma_{\langle{I}\rangle}$ is described above. The quality index
discussed above is also listed. Other columns of Table~\ref{tbl4} are
explained in the next section.
 
\section{The Period-Luminosity Relation and the Distance Modulus} 
\subsection{The P-L Diagrams in $V$ and $I$} 

As in the previous papers of this series we adopt the P-L relation in
$V$ from \cite{mad91} (1991) as

\begin{equation}
 M_{V} ~~=~~ -2.76 ~\log P - 1.40~,
\end{equation} 
whose companion relation in $I$ is
\begin{equation}
 M_{I} ~~=~~ -3.06 ~\log P - 1.81~.
\end{equation} 
The zeropoint of equations (1) and (2) is based on an adopted LMC
modulus of 18.50.

\begin{figure*}
\centerline{\hbox{\psfig{figure=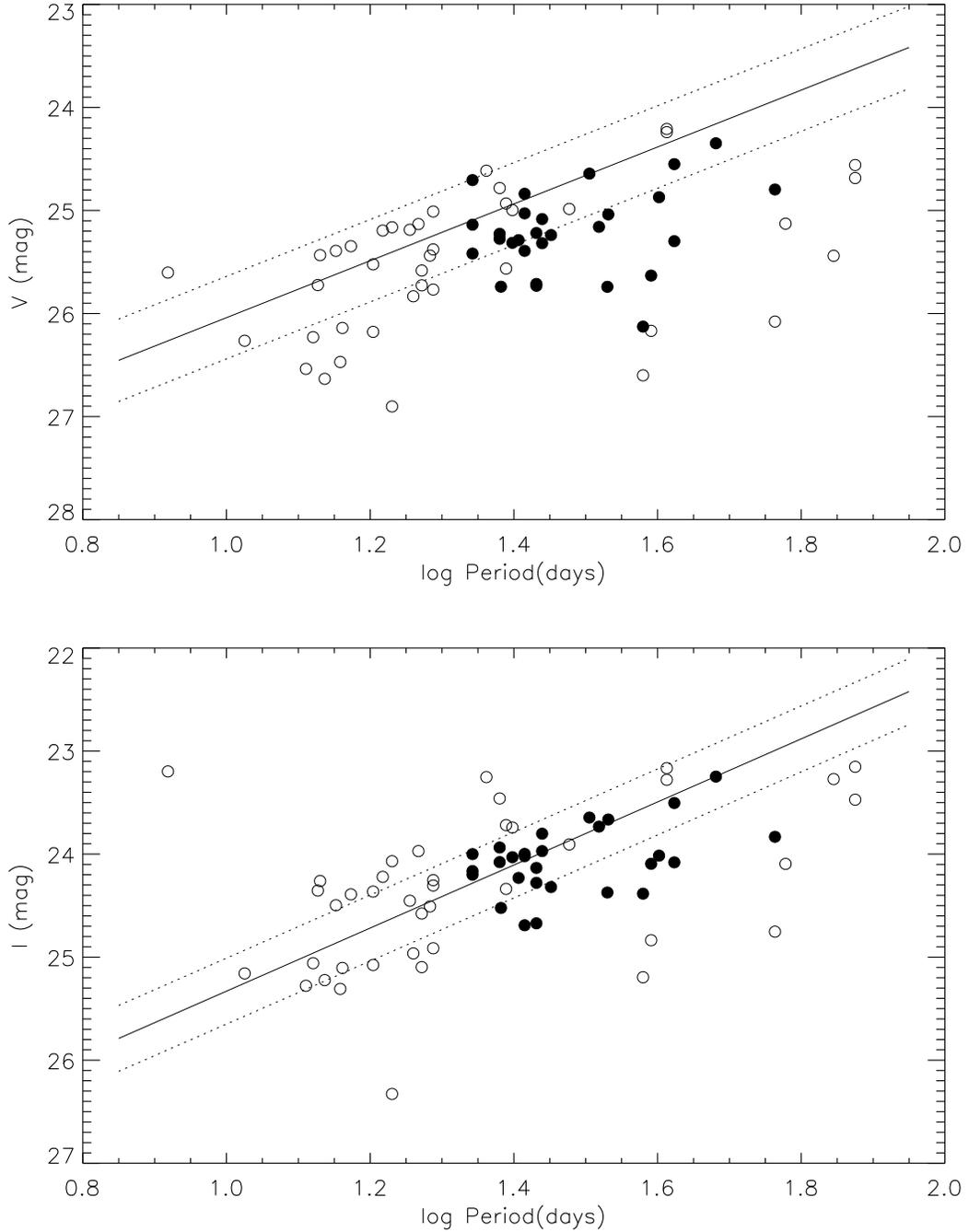,width=15.0cm}}}
\caption{The apparent P-L relations in $V$ and $I$ showing all the
Cepheid data in Table~\ref{tbl4}.  The solid circles are Cepheids with
$\log P > 1.35$ with quality class of 4 or better.  Open circles are
for the remainder of the variables in Table~\ref{tbl4}. The drawn
lines are the adopted P-L relations in equations (1) and (2) used in
the previous papers of this series from \cite{mad91} (1991).  The
ridge line has been put arbitrarily at a modulus of $(m -M) = 30.2$
simply to guide the eye before analysis of differential extinction.
\label{fig6}}
\end{figure*}

The P-L relations in $V$ and $I$ for the 68 Cepheids in
Table~\ref{tbl4} are shown in Fig.~\ref{fig6}.  The filled circles
show objects with periods between 20 and 58 days that have a quality
index of 4 or higher.  These are the best observed Cepheids.  No
selection based on color has been made here, hence the total range of
differential extinction values is contained in the data as plotted,
explaining part of the large scatter.

The continuous line in each of the two panels shows equations (1) and
(2) as the ridge-line relation using an apparent distance modulus of
30.2. The faint dashed upper and lower envelope lines indicate the
expected scatter about the mean due to the intrinsic width of the
Cepheid instability strip in the HR diagram (\cite{sata68} 1968).

The large observed scatter of the data outside these envelope lines
are due to the combination of (1) measuring and systematic errors due
to background and contamination, (2) the random error of photon
statistics, and (3) the large effects of the variable extinction
evident from Fig.~\ref{fig2}.  Any modulus inferred directly from
Fig.~\ref{fig6} would be unreliable.  No matter how the lines defining
the P-L strip are shifted vertically in the two panels of
Fig.~\ref{fig6}, a large fraction of the points of the total sample
will remain outside the boundaries of the instability strip.  Analysis
of the scatter is the subject of the next two subsections.

\subsection{Deriving the Distance Modulus}
\subsubsection{A Preliminary Analysis of the P-L Relation} 

Inspection of Fig.~\ref{fig6} reveals that the deviations from the
ridge-lines in $V$ and $I$ of individual Cepheids are correlated.
Stars that deviate faintward in $V$ also generally deviate faintward
in $I$ and vice versa.  This, of course, is a signature of variable
extinction, but it can also be caused if the {\it systematic}
measuring errors are correlated due, for example, to confusion or to
errors in the compensation for background contamination that are not
independent in the $V$ and $I$ passbands.  To explore these
possibilities and to correct for them we use the tools developed in
Paper~V and used again in Papers~VII and VIII.

For each Cepheid we calculate the apparent distance moduli separately
in $V$ and in $I$ from the P-L relations of equations (1) and (2) and
the observed $\langle{V}\rangle$ and $\langle{I}\rangle$ magnitudes
from Table~\ref{tbl4}. These apparent distance moduli, called $U_{V}$
and $U_{I}$ in columns (7) and (8) of Table~\ref{tbl4}, are calculated
by
\begin{equation}
 U_{V} ~~=~~ 2.76 ~\log P + 1.40 + \langle{V}\rangle~,
\end{equation}
and
\begin{equation}
 U_{I} ~~=~~ 3.06 \log P + 1.81 +\langle{I}\rangle~.
\end{equation} 
They are the same as equations (6) and (7) of Paper~V.

\begin{figure*}
\centerline{\hbox{\psfig{figure=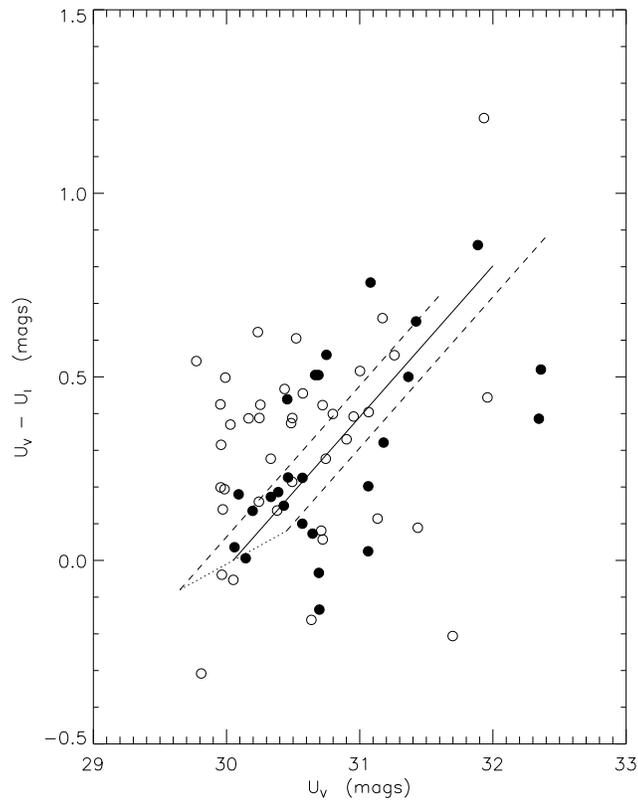,width=9.0cm}}}
\caption{Diagnostic diagram for the detection of differential
reddening, described in the text. \label{fig7}}
\end{figure*}

If the differences between the $V$ and $I$ moduli are due solely to
reddening, and if the dependence of the reddening curve on wavelength
is the normal standard dependence as in the Galaxy, then the true
modulus $U_T$ is given by
\begin{equation}
 U_{T} ~~=~~ U_{V} - R'_{V} \cdot (U_{V} - U_{I})~,
\end{equation} 
where $R$ is the ratio of total to selective absorption,
$A_{V}/E(V-I)$.  This is equation (8) of Paper~V. However, equation
(5) is valid only if the difference between $U_{V}$ and $U_{I}$ is due
to extinction, not to correlated measuring errors, in which case the
value of $R$ would {\it not} be given by the normal extinction curve
where $A_{V}/A_{I} = 1.7$ and the ratio of absorption to reddening is
$R'_{V} = A_{V}/E(V-I) = 2.43$ (\cite{schef82} 1982). Such coupled
errors clearly exist (based on Fig.~\ref{fig7} later where the slope
of the $U_{V}$ vs. $U_{I}$ correlation is closer to 1 than to the
required slope of $A_{V}/A_{I} = 1.7$ if the correlation were to be
due entirely to differential extinction rather than to measuring
errors). Hence, the interpretation of the values calculated from
equations (3), (4), and (5) is considerably more complicated than
would be the case in the absence of the correlated systematic
measuring errors.  Nevertheless, in the initial pass at the data to
derive the true modulus, we use equations (3) to (5) as a first
approximation. The second approximation, based on knowledge gained by
the methods of this section, is in the next section 4.2.2.

The values of $U_{T}$ are listed in column 9 of Table~\ref{tbl4}.
These would be the true moduli, as corrected for normal extinction,
assuming that there are no systematic measuring errors.  The total rms
uncertainty for each $U_{T}$ value is listed in column 10.  This
uncertainty includes contributions from the estimated random measuring
errors in the mean $V$ and $I$ magnitudes, (in columns 4 and 6), as
propagated through the de-reddening procedure, as well as the
uncertainty associated with the intrinsic width of the P-L relation,
i.e. a given Cepheid may not be on the mean ridge-line of the P-L
relation. The de-reddening procedure amplifies the measuring
errors. Therefore many Cepheids are needed to beat down these large
errors (notice the very large values in column 10) in any final value
of the modulus.  The values shown in column 10 of Table~4 were
calculated using equations 9, 10 and 11 of Paper~V, and correspond to
$\sigma_{tot}^2$ as defined in Paper~V. However, note that equation
(11) in Paper~V should be :
\begin{displaymath}
 \sigma_{width}^2 = (R'_{V}-1)^2 \cdot \rho_{V}^2 + {R'_{V}}^2 \cdot
 \rho_{I}^2~.
\end{displaymath} 
Our records show that while this equation was given incorrectly in
Paper~V, the calculations were done with the correct relation.

Various arithmetics done on the $U_{T}$ values in column 9 of
Table~\ref{tbl4} give the first indication of the true modulus.
Consider first all Cepheids of all quality indices, but excluding
C1-13 and C3-V15 because of their extreme values of $U_{T}$.  The 66
Cepheids in this sample give $\langle{U_{T}}\rangle = 29.90 \pm 0.08$
(the error estimate is based on the adopted rms of a single Cepheid of
$0\fm646$).  Restricting the sample to only those Cepheids with good
to excellent data defined by a quality index of 4 or higher gives a
subsample of 41 Cepheids for which $\langle{U_{T}}\rangle = 30.09 \pm
0.085$ (error based on an rms of $0\fm535$).  Note that the 3-sigma
upper and lower limits on the true modulus from this arithmetic are
30.34 and 29.83.  These are therefore very strong upper and lower
limits on the true distance modulus of NGC~3627.

These are the best values we can derive without making cuts in the
data according to period, QI, and/or color that would select the
bluest and least reddened, and that would be least affected by
selection bias at the low-period end.  We consider now such
subsamples, first using cuts in period and QI, but not yet in color
which follows in the next section.

A plot of $U_{T}$ vs. period (not shown) reveals a trend that objects
with the shortest periods yield smaller $U_{T}$ moduli.  This result
has been seen by us in the previously analyzed galaxies, and by other
investigators analyzing similar data in yet other galaxies.  It is due
to a combination of selection bias at the short-period end
(\cite{san88} 1988) as well as non-symmetrical observational bias in
measuring colors near the faint limit, skewing $U_{T}$ via the color
effect in equation (5).  We note that this trend disappears once
periods are restricted to longer than 25 days. We also note that the
variation of $U_{T}$ with Quality Index is not as acute for these data
as we have seen in previous cases.  A likely reason is that $I$
magnitudes were obtained at five epochs, while in all except one of
our previous papers fewer epochs were available.  At any rate there is
no compelling trend in $U_{T}$ once the sample is restricted to
objects with QI $\geq 3$.

Using the sample of 27 Cepheids that have periods $\geq 25$ days but
shorter than the baseline of 58 days, and that have QI $\geq 3$, and
weighting the individual $U_{T}$ values by $(1/{\rm rms})^2$, gives
the mean de-reddened modulus of $(m - M)_0 = 30.04 \pm 0.12$.  If we
make more restrictive cuts by accepting only objects with QI $\geq 4$,
and then $\geq 5$, and then 6, (with the same period cuts as above),
we obtain respectively weighted mean ``true'' moduli of $29.99 \pm
0.13$, $30.10 \pm 0.15$, and $30.07 \pm 0.18$.  This shows the general
stability of the result.  For the unweighted average of the 27
Cepheids we obtain $(m - M)_0 = 30.10 \pm 0.14$.  Due to the fact that
the P-L relations in $V$ and $I$ have non-negligible width at a
constant period, a Cepheid that is {\it intrinsically} redder and
fainter will, on the average, carry {\it larger} measuring
errors. This can contribute to systematically underestimating the
distance when individual $U_{T}$'s are weighted as above.  The
unweighted solution is therefore preferred, since the uncertainties
are similar.

We note again that the $U_{T}$ values so derived depend on the
assumption that the differences between $U_{V}$ and $U_{I}$ are due to
reddening alone, in the absence of appreciable systematic and
correlated measuring errors, or when the errors for $U_{V} - U_{I}$
{\it are distributed symmetrically}.  If equation (5) is used for
Cepheids {\it where correlated and/or asymmetrical errors in $V$ and
$I$ dominate over differential reddening}, thereby producing a ratio
of the $V$-to-$I$ errors that is different from 2.43, the $U_{T}$
derived via equation (5) {\it will be in error}.

In particular, several Cepheids which were discovered in $V$ are too
faint in $I$ to be measured, as already mentioned.  This introduces a
selection effect that biases against Cepheids with bluer colors.  The
effect is most pronounced at short periods where the {\it intrinsic}
colors are bluest.  This effect gives an asymmetrical distribution of
errors in $U_{V} - U_{I}$ in the sense that it makes the de-reddened
modulus {\it too small}.

In Paper~V we devised a method to test for the presence of
differential extinction or for the fact that the scatter about the P-L
relation is due predominantly to measuring errors, or a combination of
both.  The method, shown in Fig.~11 of Paper~V for NGC~4536 and
explained in the Appendix there, was used in Paper~VI for NGC~4496A
(Fig.~9 there) and in Paper~VIII for NGC~4639 (Fig.~7 there).  The
method is to plot the difference in the apparent$V$ and $I$ moduli for
any given Cepheid as ordinate against the apparent $V$ modulus as
abscissa.  If there is a systematic trend of the data along a line of
slope $dU_{V}/d(U_{V} - U_{I}) = 2.43$, then equation (5) applies and
there is clearly differential reddening.  If, on the other hand, there
is a general scatter with no trend, that scatter is dominated by
measuring errors. While in the latter case true differential
extinction can be hidden by measuring errors, trying to correct for
putative reddening will result in interpreting any asymmetry in the
error distribution as specious extinction.  In two of the three
previous cases, NGC~4536 (Paper~V), and NGC~4496A (Paper~VI) there is
no trend along a differential reddening line.  In the third case of
NGC~4639 (Paper~VIII), there is a slight trend but also large scatter
showing that the spread of points appears to be due to a mixture of
measurement errors as well as from differential extinction.

The diagnostic diagram just described is shown for the NGC~3627 data
from Table~\ref{tbl4} in Fig.~\ref{fig7}.  The filled circles show
Cepheids with periods between 25 and 58 days. The solid line indicates
the reddening vector for the P-L ridge line, if the true (de-reddened)
distance modulus is 30.05 which is close to the mean derived earlier
in this section.  The dashed lines show the bounds due to the
intrinsic dispersion of the P-L relation as explained in Paper~V.  The
slope of the lines is $A_{V}/E(V-I) = 2.43$ as before.

There is only marginal evidence from Fig.~\ref{fig7} for a general
trend along the solid line.  The spread of points clearly spills
outside these bounds, indicating that a very significant fraction of
the scatter is due to measuring errors and related biasses.  If such
errors are distributed symmetrically, equation (5) will yield the
correct answer, but if there are correlated errors in $V$ and $I$, or
if there are selection effects that depend on color, using equation
(5) will introduce errors.  Note that the scatter of points is skewed
along a direction orthogonal to the reddening vector: the lower right
side appears to be more sparsely filled, indicating a possible
selection effect.  A more detailed inspection of the spread in
Fig.~\ref{fig7} shows that the largest scatter occurs in the reddest
Cepheids, using the $(\langle{V}\rangle - \langle{I}\rangle)$ colors
computed from columns (3) and (5) of Table~\ref{tbl4}. This implies
that the reddest Cepheids are so more because of skewed measurement
errors than due to bona-fide reddening. This is not to deny the
presence of differential reddening, but an acknowledgement that
equation (5) alone is not adequate for obtaining a bias free
result. We proceed by making an additional restriction of the data by
color. Note that such a restriction used in conjunction with equation
(5) does not introduce a procedural bias in the distance modulus.

\subsubsection{The Distance Modulus By Restricting The Data By Color} 
 
A plot (not shown) of the color-period relation from the data in
Table~\ref{tbl4} shows a distinctive separation into two major color
groups, one close to the intrinsic $({\langle{V}\rangle}_0 -
{\langle{I}\rangle}_0)$-period relation known for unreddened Cepheids
in the Galaxy, LMC, and SMC as summarized in \cite{sanetal99} (1999)
from data by \cite{dea78} (1978), \cite{cal85} (1985), and
\cite{fern90} (1990).  The other group of Cepheids with
$(\langle{V}\rangle - \langle{I}\rangle)$ colors larger than 1.15 are
far removed from the intrinsic domain in the color-period plot.  They
are also the Cepheids that show the largest deviation faintward in the
P-L relations of Fig.~\ref{fig6}.  Excluding these as the Cepheids
with the largest reddening leaves a subsample of 29 Cepheids (the blue
group) with $QI \geq 3$, and with $1.15 < \log P < 1.76$.  We also
have excluded C1-V13 because it is obviously an outlier.  C1-V1,
although blue, is excluded because its proposed period of 75 days is
outside the baseline of 58 days.

\begin{figure*}
\centerline{\hbox{\psfig{figure=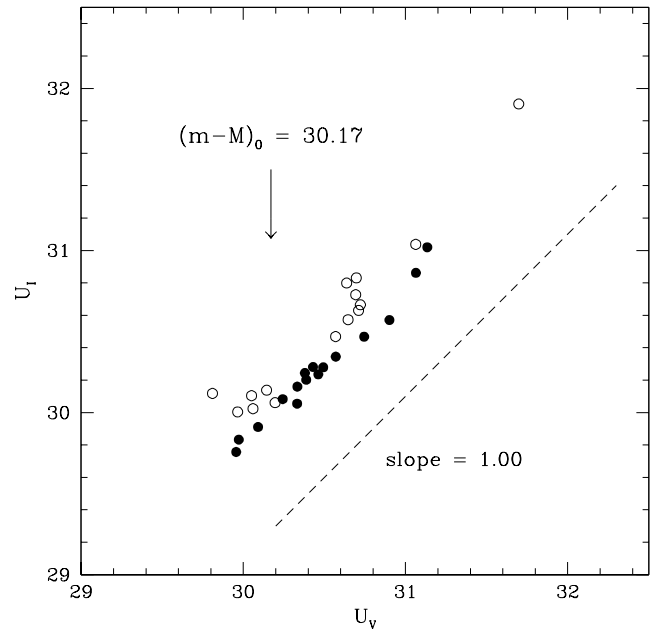,width=9.0cm}}}
\caption{Correlation of the apparent modulus in $V$ with that in $I$
for the subset of the Cepheids in Table~\ref{tbl4} with
$(\langle{V}\rangle - \langle{I}\rangle) < 1.15$ and $\log P >
1.15$. Open circles are the bluest of the 29 Cepheid subsample. Closed
circles are of intermediate color. \label{fig8}}
\end{figure*}

Figure~\ref{fig8} shows again that differential reddening is not the
major factor in the scatter of the P-L relation, where $U_{V}$ is
plotted vs. $U_{I}$ for the 29 blue Cepheids of the subsample.

There is a clear correlation of $U_{V}$ and $U_{I}$, but the slope is
not $A_{V}/A_{I} = 1.7$ as required if the Cepheids below the ridge
lines of Fig.~\ref{fig6} were fainter because of a larger differential
extinction.  Rather, the slope near 1 in Fig.~\ref{fig8} can only be
due to {\it correlated measuring errors} as we suspected in the last
section.  Of course, a part of the correlation must also be due to
reddening, if for no other reason than the obvious dust pattern in
Fig.~\ref{fig1} and \ref{fig2}.

\begin{figure*}
\centerline{\hbox{\psfig{figure=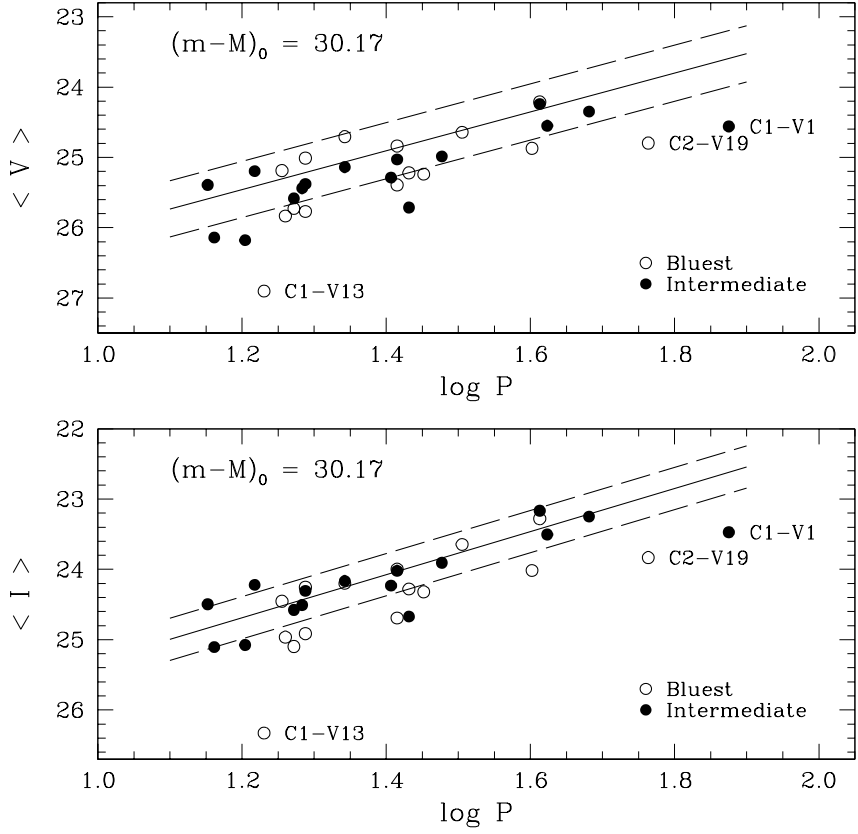,width=12.0cm}}}
\caption{The P-L relations for the subset of Cepheids in
Table~\ref{tbl4} with $(\langle{V}\rangle - \langle{I}\rangle) < 1.15$
and $\log P > 1.15$.  Same coding as in Fig.~\ref{fig8}. \label{fig9}}
\end{figure*}

Figure~\ref{fig9} shows the P-L relations for the subset of the 29
bluest Cepheids.  The scatter is markedly reduced from that in
Fig.~\ref{fig6}, showing that the color cut has produced a subset with
the smallest extinction and/or measuring error.

We use the $U_{V}$ and $U_{I}$ apparent moduli in Table~\ref{tbl4}
calculated from equations (3) and (4) of \S~4.2.1, and calculate mean
values $\langle{U_{V}}\rangle$ and $\langle{U_{I}}\rangle$ using the
29 Cepheids of this subset.  Assuming that the measuring errors for
this sample are random and that they cancel in the mean permits the
premise that the {\it difference} in the mean apparent moduli in the
$V$ and $I$ passbands {\it is} now due to reddening.  Multiplying the
mean modulus difference by $A_{V}/E(V-I) = 2.43$ then gives $A_V$,
which when subtracted from $\langle{U_{V}}\rangle$ gives the true
modulus.  This, of course, is what equation (5) does automatically,
hence we need only analyze the $U_{T}$ values in Table~\ref{tbl4} for
the 29 Cepheid subsample.

\begin{figure*}
\centerline{\hbox{\psfig{figure=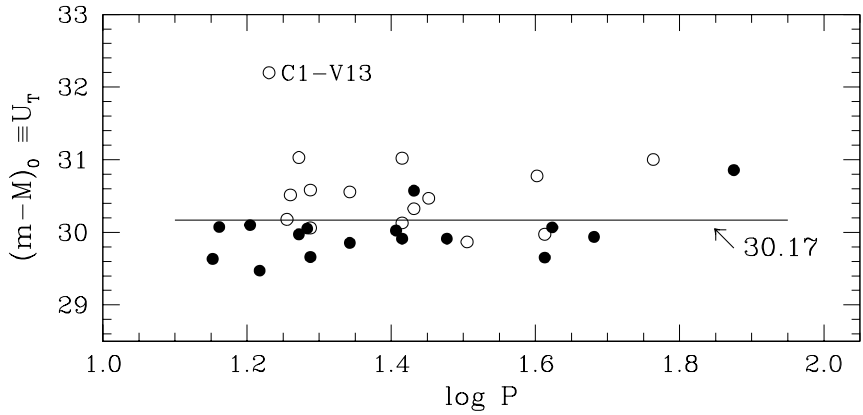,width=12.0cm}}}
\caption{$U_T$ vs. $\log P$ for the subset of the 29 bluest
Cepheids.  Same coding as in Figs.~\ref{fig8} and \ref{fig9}.
\label{fig10}}
\end{figure*}

The weighted mean $\langle{U_{T}}\rangle_{W}$ for the subset of 29
Cepheids gives $\langle{U_{T}}\rangle = 30.12 \pm 0.11$.  The
individual $U_{T}$ values are plotted vs. $\log P$ in
Fig.~\ref{fig10}.  As noted in the last section and seen in
Fig.~\ref{fig10}, there is a tendency for the shortest period Cepheids
to have the smallest individual moduli.  Making the period cut at
$\log P > 1.25$ removes the tendency and gives
\begin{equation}
   {\langle{U_{T}}\rangle}_W = 30.17 \pm 0.12~,
\end{equation} 
as the weighted mean from the 25 Cepheids with $1.25 < \log P < 1.78$,
which we adopt. The unweighted mean is $\langle{U_{T}}\rangle = 30.24
\pm 0.09$.

\begin{figure*}
\centerline{\hbox{\psfig{figure=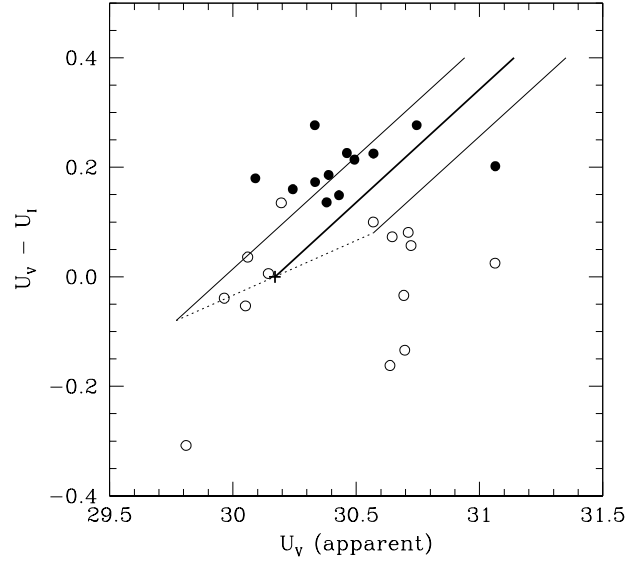,height=10.0cm}}}
\caption{Same as Fig.~\ref{fig7} but for the blue 25
Cepheid subsample.  The zero-point of the correlation is assumed to be
at $(m - M)_0 = U_T = 30.17$ as in equation (6). \label{fig11}}
\end{figure*}

Justification for our restriction to the subsample of 25 Cepheids is
given in Fig.~\ref{fig11} which is the diagnostic diagram of
Fig.~\ref{fig7} but using only this subsample.  Plotted are again the
individual apparent moduli $U_{V}$ vs.  the difference between the
individual $V$ and $I$ apparent moduli.

Figure~\ref{fig11} is much cleaner than Fig.~\ref{fig7}, and the data
points now scatter nearly symmetrically about the differential
reddening line.

Applying, as in previous papers of this series, the correction for the
``long'' vs. ``short'' exposure effect of $0\fm05$ to equation (6),
the de-reddened modulus of NGC~3627 is
\begin{equation}
   (m - M)_0 = 30.22 \pm 0.12~,
\end{equation} 
which we adopt.

\section{The Absolute Magnitude at Maximum of SN~1989B Added to
   Previous Calibrators; the Present Status of the Calibration}

The light curves in the {\it UBVRI} passbands of SN~1989B are well
determined near maximum light, giving $B_{\rm max} = 12.34 \pm 0.05$,
and $V_{\rm max} = 11.99 \pm 0.05$ (\cite{wel94} 1994).  These authors
have also determined the reddening of SN~1989B itself to be $E(B-V) =
0.37 \pm 0.03$.  Hence, the de-reddened magnitudes at maximum light
are $B_{\rm max}^{0} = 10.86 \pm 0.13$ and $V_{\rm max}^{0}= 10.88 \pm
0.10$.  The absolute magnitudes at maximum for SN~1989B are
\begin{equation}
   M^0_B({\rm max}) = -19.36 \pm 0.18~,
\end{equation} 
and
\begin{equation}
   M^0_V({\rm max}) = -19.34 \pm 0.16~,
\end{equation} 
based on equation (7) and on the Cepheid zero points of equations (1)
and (2).

These values are combined in Table~\ref{tbl5} with our previous
calibrations of the six SNe~Ia from Papers I--VII of this series.  In
addition, two new values are included from data for SN~1974G in
NGC~4414 (\cite{tur98} 1998; \cite{schae98} 1998), and SN~1998bu in
NGC~3368 (M~96) whose distance modulus is by \cite{tan95} (1995) with
photometry of the SN reported by \cite{sun98} (1998). All Cepheid
distances from the WFPC2 are corrected by $0\fm05$ for the short
vs. long exposure photometric zeropoint difference (\cite{ste95} 1995;
\cite{saha96a} 1996a).

Neglecting the listed $M_B$ of SN~1895B, which is the most uncertain
of the group, and because it is not absolutely certain whether it was
spectroscopically normal at all phases, gives the straight mean values
of $\langle{M_B}\rangle = -19.49 \pm 0.03$ and $\langle{M_V}\rangle =
-19.49 \pm 0.03$.  We adopt the {\it weighted} means, giving the mean
calibration without any second parameter corrections for decay rate
(see \S~7) as
\begin{equation}
   \langle{M_B(\rm max)}\rangle = -19.49 \pm 0.07~,
\end{equation} 
and
\begin{equation}
   \langle{M_V(\rm max)}\rangle = -19.48 \pm 0.07
\end{equation} 
on the Cepheid distance scale of equations (1) and (2).  Inclusion of
SN~1895B would give $\langle{M_B}\rangle = -19.54 \pm 0.06$ for the
straight mean and $\langle{M_B}\rangle = -19.52 \pm 0.07$ for the
weighted mean.

  We have been criticized for using calibrators such as SN~1960F and
SN~1974G, for which the photometry is more uncertain than others. We
should point out that the net worth is not just the uncertainty in the
photometry of the supernova, but the combined uncertainty with that of
the Cepheid distance determination. Given the range of uncertainties
in the Cepheid distance determinations (which depend on
distance/faintness, crowding, etc), even photometric uncertainties
considerably worse than for SN~1960F can be tolerated if the Cepheid
determinations are as good as they are for its host galaxy
NGC~4496A. In addition, we weight the contribution of individual
calibrators by the inverse variance from the combined uncertainty of
distance and supernova photometry, a procedure consistent with
Bayesian inference.  Thus even bona-fide `poor' cases like SN~1974G
(which has also a relatively poor Cepheid distance to its host galaxy
NGC~4414), enter only with appropriately lowered weight.

Equations (10) and (11), based now on eight calibrators (neglecting
SN~1895B), are similar to equations (12) and (13) of Paper~VIII
(\cite{saha97} 1997) based there on six calibrators (again neglecting
SN~1895B).  Equation (10) here is $0\fm03$ fainter than in Paper~VIII.
Equation (11) here is identical with that of Paper~VIII.  Because of
the similarity of the equations (10) and (11) here with equations (12)
and (13) in Paper~VIII, the interim value of the Hubble constant, sans
decay-rate corrections, for this stage of our $HST$ experiment is
nearly identical with the values set out in Table~7 of Paper~VIII.
The means of all values in that table, depending on how the SNe~Ia
Hubble diagram is divided between {\it spirals} observed before and
after 1985, and/or with redshifts larger or smaller than $\log v_{220}
= 3.8$, and using equation (10) here rather than equation (12) of
Paper~VIII, are then
\begin{equation}
   \langle{H_0(B)}\rangle = 58 \pm 2~~({\rm internal})\ksm~,
\end{equation}
and
\begin{equation}
   \langle{H_0(V)}\rangle =59 \pm 2~~({\rm internal})\ksm~.
\end{equation}

\section{The Hubble Diagram of SNe~Ia and the Value of $H_0$
if Second Parameter Corrections are Made}

The Hubble diagram for blue SNe~Ia with $(B_{\rm max}-V_{\rm max}) <
0.20$ , based on many sources, historical as well as modern, was given
in Paper~VIII (\cite{saha97} 1997, Fig.~10).  When calibrated with the
SNe absolute magnitudes, this diagram gives the Hubble constant
directly.  The details of the sample are being published in a paper by
\cite{par99} (1999), and we make use of these data in this section to
calibrate again correlations of absolute magnitudes of SNe~Ia with the
often adopted second parameters of decay rate of the light curve,
color of the SNe at maximum, and galaxy type.

The precept adopted in the early papers of this series was that by
restricting both our calibrator SNe~Ia and the Hubble diagram of
SNe~Ia to ``Branch normal'' events (\cite{bra93} 1993), we have
selected a homogeneous sample of SNe~Ia for which any systematic
variation of absolute magnitude among the sample will be small enough
to be neglected to first order at the 10\% level (\cite{cad85} 1985;
\cite{lei91b} 1991b; \cite{sata93} 1993; \cite{tasa95} 1995). In the
meantime, thanks to the observational program of the Calan/Tololo
Chilean consortium (\cite{ham95} 1995; 1996a, b) and the theoretical
insights of variations in the pre-explosion conditions of the
progenitors (cf. \cite{vHipp97} 1997, \cite{hof98} 1998, \cite{nad98}
1998 for summaries), it has become clear that there is in fact a {\it
continuous} variation of SNe~Ia properties, including absolute
magnitude, that can be detected from observed second- parameter
properties and which can now be accounted for even at this $< 10\%$
level.

The most apparent of these second parameters is the change in the
shape of the light curve, quantified by the rate of decay from maximum
light (\cite{rus74} 1974; \cite{dev76} 1976; \cite{psk77} 1977, 1984;
\cite{phi87} 1987; \cite{bar90} 1990; \cite{bra92} 1992; Hamuy et al.\
1996a, b).  \cite{phi93} (1993) derived a very steep relation between
decay rate and absolute magnitude using distances determined by a
variety of methods and including also very red SNe~Ia.  A much flatter
dependence was found by \cite{tasa95} (1995) for blue SNe~Ia using
more reliable relative distances from recession velocities.  The
flatter slope was confirmed for only the blue SNe~Ia by the subsequent
extensive Calan/Tololo data (Hamuy et al.\ 1996a, b; \cite{saha97}
1997).
         
A detailed discussion of the second order corrections to the SNe~Ia
distance scale depending on second and third parameter correlations is
the subject of the accompanying paper by \cite{par99} (1999) where it
is shown that besides the decay rate, color at maximum is also a
principal second parameter, confirming a similar result by
\cite{tri98} (1998).
 
The Parodi et al.\ paper lists relative kinematic absolute magnitudes
as if the local Hubble redshift-to-distance ratio is the same for
redshifts smaller than $10,000\kms$ as for the remote $H_0$ for $v >
10,000\kms$, neglecting the evidence set out in Paper~VIII
(\cite{saha97} 1997, \S~8; cf. also \cite{zeh98} 1998) that the local
value may be 5\% to 10\% larger than the global value.

This possible change of the Hubble ratio outward, decreasing with
distance by $\leq 10\%$ for $v$ out to $ 10,000\kms$, is also
consistent with other external data on first ranked cluster galaxies
(\cite{lau94} 1994; \cite{tam98a} 1998a).  The suggestion is also
consistent with the derived shallow slope of $d\log v/dm = 0.192$ of
the local ($v < 10,000\kms$) Hubble diagram by Parodi et al.\ (1999,
their eqs. 7 and 8) rather than 0.200 required if $H_0$ did not vary
with distance.

In the analysis of this section we consider the consequences of a
variable $H_0$ decreasing outward, that is required to give a slope of
0.192 to the Hubble diagram locally, and then carry again the analysis
of the second parameter effects of decay rate, color, and galaxy type
using the derived kinematic absolute magnitudes.
\subsection{Kinematic Absolute Magnitudes Using a Variable Hubble 
Constant with Local Distance for $v < 10,000\kms$ }
We accept the premise that the slope of the Hubble diagram for
redshifts smaller than $10,000\kms$ is 0.192 (\cite{par99} 1999).  If
we adopt an arbitrary global (remote field) value of $H_0 = 55$ (to be
adjusted later by our calibrators in Table~\ref{tbl5}), then, by an
obvious calculation, the variation of $H_0$ in the distance interval
of $1000 < v < 10,000$ is well approximated (within $1\%$) by
\begin{equation}
   H_0(v) = -5.39 \log v + 76.50~.
\end{equation}
This gives $H_0 = 60.3$ at $v = 1000\kms$ and $H_0 = 54.9$ at $v =
10,000\kms$.

Equation (14) has been used to recalculate the absolute magnitudes of
all SNe~Ia in the fiducial sample in Table~1 of \cite{par99} (1999)
for which $\log v < 4.00$. $H_o = 55$ has been assumed for $\log v >
4.00$.  The results are set out in Table~\ref{tbl6} which is divided
into three parts according to Hubble type to better understand the
type dependence of the Hubble diagram seen in Fig.~10 of Paper~VIII
(\cite{saha97} 1997).  The first section of Table~\ref{tbl6} lists the
SNe in late type spirals ($T$ of 3 and greater, meaning Sb to Im
types).  The second section for $T = 1$ and 2 are for Sa and Sab
parent galaxies. The third section lists parent galaxy $T$ types of 0
and smaller (E and S0).

Columns (1) through (6) repeat data from Table~1 of \cite{par99}
(1999) with magnitudes at the respective maxima denoted by $B_0$,
$V_0$, and $I_0$.  The redshifts in column (3) are corrected for
peculiar motions.  For $v < 3000\kms$ the redshifts, reduced to the
frame of the centroid of the Local Group (\cite{yah77} 1977), were
then corrected again to the frame of the Virgo cluster using the
self-consistent Virgocentric infall model with the local infall vector
(actually retarded expansion of the Local Group relative to Virgo) of
$220\kms$ (\cite{kra86ab} 1986a, b).  For $v > 3000\kms$ an additional
correction of $630\kms$ relative to the CMB frame due to the CMB
dipole anisotropy (\cite{bou81} 1981; \cite{wil84} 1984) was applied
according to the model of \cite{tasa85} (1985, their Fig.~2).  For
more distant galaxies we have adopted the corrected velocities from
\cite{ham96a} (1996a).  The sources for the photometry in columns (4)
to (6) are listed in \cite{par99} (1999).  The magnitudes are
corrected for Galactic absorption (\cite{bur84} 1984).

Column (7) is the adopted local value of $H_0$ calculated from
equation (14) for $\log v < 4.00$, and using $H_0 = 55$ for larger
redshifts.  The resulting distance moduli are in column (8).  The
corresponding absolute magnitudes are in columns (9) to (11).  The
decay rates $\Delta m_{15}(B)$ in column (12) are from the sources
listed by \cite{par99} (1999).

The data for the eight calibrators in Table~\ref{tbl5}, except for the
observed apparent magnitudes, are not shown in Table~\ref{tbl6}
because their absolute distance moduli are on the Cepheid system, not
based on redshifts.

The means of columns (9) to (11) for the absolute magnitudes in $B$,
$V$, and $I$ are shown at the foot of each of the three sections of
Table~\ref{tbl6}.  The systematic progression, becoming fainter for
the earlier galaxy types, is evident and is significant at the 2-sigma
level.  It is this difference that causes the separation of the ridge
lines in the Hubble diagram of spirals and E and S0 galaxies seen in
Fig.~10 of \cite{saha97} (1997).

However, this is not a type dependence per se.  Note in column 12 that
the mean decay rates, shown at the foot of each section, differ
significantly between the three sections of Table~\ref{tbl6}.  The
decay rates are much longer for the E to S0 types, averaging
$\langle{\Delta m_{15}(B)}\rangle = 1.44 \pm 0.04$ for these early
Hubble types, compared with $\langle{\Delta m_{15}(B)}\rangle = 1.03
\pm 0.03$ for the spirals in the first section of Table~\ref{tbl6}.

Because the decay rate itself is correlated with absolute magnitude
(next section), the {\it apparent} dependence of $\langle{M({\rm
max})}\rangle$ on Hubble type is in fact due to the decay rate
correlation.  This, of course, is a clue as to differences in the
progenitor mass of the pre-SNe~Ia as a function of Hubble type, and
goes to the heart of the physics of the phenomenon (eg. \cite{nad98}
1998).  In any case, the apparent correlation of $\langle{M}\rangle$
with Hubble type disappears when the decay rate corrections of the
next section are applied.

Said differently, the {\it apparent} dependence of mean absolute
magnitude with Hubble type is due to the difference in mean decay rate
between early type galaxies (E to S0 types) and late type spirals (the
first section of Table~\ref{tbl6}), together with the dependence of
decay rate on absolute magnitude at maximum (Fig.~\ref{fig12} in the
next section).

\subsection{The Decay Rate Dependence}

\begin{figure*}
\centerline{\hbox{\psfig{figure=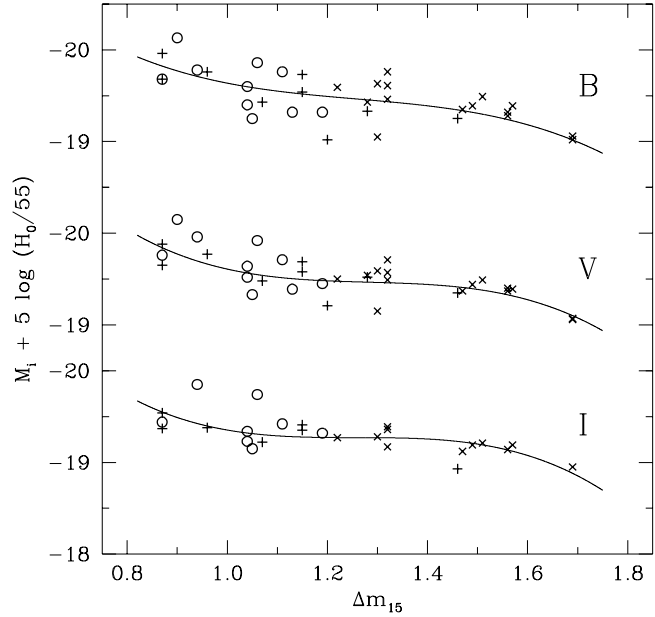,height=10.0cm}}}
\caption{Correlation of the kinematic absolute magnitudes $M_i$
($H_0 = 55$) from columns (9) -- (11) of Table~\ref{tbl6} with the $\Delta
m_{15}(B)$ decay rate from column (12).  Open circles are for late
type spirals, Roman crosses are for early type spirals, and
skipping-jack crosses are for E and S0 galaxies. The lines are from
equations (15) -- (17) with the adopted zero points at $\Delta m_{15}(B)
= 0$ that are listed in the text. \label{fig12}}
\end{figure*}

The absolute magnitudes in columns (9) to (11) of Table~\ref{tbl6} are
plotted in Fig.~\ref{fig12}.  It is clear that there is a relation of
decay rate with (kinematic) absolute magnitude (\cite{phi93} 1993),
but that the relation is strongly non-linear.  There is a clear
plateau between abscissa values of decay rates that are greater and
smaller than 1.0 to 1.5. In this plateau region, the absolute
magnitude is nearly independent of the decay rate.

Figure~\ref{fig12} shows the correlations of columns (9), (10), and
(11) (of Table~6) with the decay rate in column (12). The correlation
with absolute magnitude for small and for large $\Delta m_{15}(B)$ is
obvious.  Nevertheless, the correlation is clearly non-linear.  There
is a central region ($1.0 < \Delta m_{15}(B) < 1.5$) where the
correlation is weak.  This is the core of the ``Branch normal''
majority of SNe~Ia, comprising 95\% of the observed SNe~Ia (only 1 in
20 of the local SNe~Ia discoveries are ``Branch abnormal''; see
\cite{bra93} 1993).  It is only when the few known abnormal SNe~Ia are
added to the ``Branch normal'', i.e. blue SNe~Ia, that they show a
wide variation in luminosity.

Consider now the apparent correlation of $\langle{M_{\rm max}}\rangle$
with Hubble type, shown in Fig.~10 of \cite{saha97} (1997,
Paper~VIII).  It is of central physical interest how the Hubble type
of the parent galaxy can produce slow-decay rate SNe~Ia in E galaxies,
and faster-decay rate SNe~Ia in later-type galaxies.  One supposes
that the mass of the pre-SNe~Ia stars must be a function of chemical
evolution in any given parent galaxy, causing the change in the
character of the SNe~Ia explosion with Hubble type (\cite{nad98}
1998).  Nevertheless, this is an astrophysical problem, not an
astronomical problem of how to use the data to determine reliable
distances.  It suffices here to note that the type dependence on
$\langle{M_{\rm max}}\rangle$ disappears when the correction, now
described for decay rate (Fig.~\ref{fig12}), is applied.

The correlation of decay rate $\Delta m_{15}(B)$ with absolute
magnitudes in $B$, $V$, and $I$ in Fig.~\ref{fig12} is striking,
confirming Pskovskii's (1977, 1984) initial suggestions.  But again,
the correlations are much less steep than suggested by \cite{phi93}
(1993) and by Hamuy et al.\ (1996b, their Fig.~2), especially in the
region of the ``Branch normal'' SNe~Ia with $0.95 < \Delta m_{15}(B) <
1.3$ where there is no strong correlation.

\cite{ham96a} (1994a) distinguish between the slope of the decay-rate
correlation with $M({\rm max})$ obtained from using TF, PN and SBF
distances to nearby SNe~Ia (as originally done by \cite{phi93} 1993),
versus that from using relative distances from redshifts at
intermediate distances. The latter sample gives a shallower slope than
that of Phillips. By restricting to a Branch normal sample, the slope
is further reduced.

In Fig.~\ref{fig12} we have fitted cubic polynomials to the data
contained in Table~\ref{tbl6} for the correlations of absolute
magnitudes in $B$, $V$, and $I$ in columns (9) to (11) with $\Delta
m(B)_{15}$ in column (12).  Denote the correction to $M_i$ for decay
rate by $y_i$, such that the ``corrected'' absolute magnitude, reduced
to $\Delta m_{15}(B) = 1.1$, is defined by
\begin{displaymath}
   M_i^{15} = M_i(~{\rm{Table~\ref{tbl6}}}) - y_i~.
\end{displaymath}

The correlations in Fig.~\ref{fig12}, based on the decay-rate data and
absolute magnitude data in Table~\ref{tbl6}, give the following cubic
corrections for decay rate vs. absolute magnitude in $B$, $V$, and
$I$.  The lines drawn in Fig.~\ref{fig12} are calculated from these
equations.  We reduce all data to $\Delta m_{15}(B) = 1.1$, and
therefore define $x = \Delta m_{15}(B) - 1.1$.
\begin{eqnarray}
y_B = 0.693x - 1.440x^2 + 3.045x^3\,, \\ y_V = 0.596x - 2.457x^2 +
4.493x^3\,, \\ y_I = 0.360x - 2.246x^2 + 4.764x^3\,.
\end{eqnarray}
Note that these polynomials are not well constrained outside the range
of $\Delta m(B)_{15}$ spanned by the SNe~Ia in this sample. In
particular, the upturn for very slow declining SNe~Ia is
uncertain. However, the cubic characterization given here is adequate
for the present arguments and calculations: it is not used in a region
where it is ill constrained.

The corrections for decay rate, $y_i$, from equations (15) to (17),
have been applied to columns (9) to (11) of Table~\ref{tbl6} to obtain
absolute magnitudes freed from the $\Delta m(B)_{15}$ parameter
effect.  The arithmetic is not shown but the mean corrected magnitudes
$\langle{M_i^{15}}\rangle$, still binned into the three morphological
groups of Table~\ref{tbl6} are listed in the first four lines of
Table~\ref{tbl7}. The corresponding mean magnitudes
$\langle{M_B^{15}}\rangle$ and $\langle{M_V^{15}}\rangle$ of the eight
calibrators from Table~\ref{tbl5} are shown in the last line. No mean
value$\langle{M_I^{15}}\rangle$row is shown in the last line because
too few calibrators have known $I_{\rm max}$.

Table~\ref{tbl7} permits two conclusions:

(1) The type dependence shown in Fig.~10 of Paper~VIII between E
galaxies and spirals has now disappeared to within one sigma
differences after the $\Delta m_{15}$ corrections are applied.

(2) Comparing the $\langle{M_B^{15}}\rangle$ and
$\langle{M_V^{15}}\rangle$ values in the fourth line (the total
sample) with the corrected values of the calibrators in the fifth line
shows the statistical difference of $0\fm14 \pm 0\fm058$ in $B$ and
$0\fm15 \pm 0\fm076$ in $V$ between the fiducial sample and the
Table~\ref{tbl5} calibrators. Because the absolute magnitudes of the
fiducial sample are based on $H_0 = 55$, these data, corrected for
decay rate, give

\begin{equation}
H_0(B)(~{\rm decay~rate}) = 58.8 \pm 2\ksm\,,
\end{equation}
and
\begin{equation}
H_0(V)(~{\rm decay~rate}) = 59.1 \pm 2\ksm\,.
\end{equation}

It is shown in \cite{par99} (1999), that the effect of the decline
rate correction taken alone, is to increase $H_0$ by 7\%. This
correction depends on the extent to which the decline rate
distribution is different for the calibrating and distant samples of
SNe~Ia.

\subsection{The Color Dependence}

\begin{figure*}
\centerline{\hbox{\psfig{figure=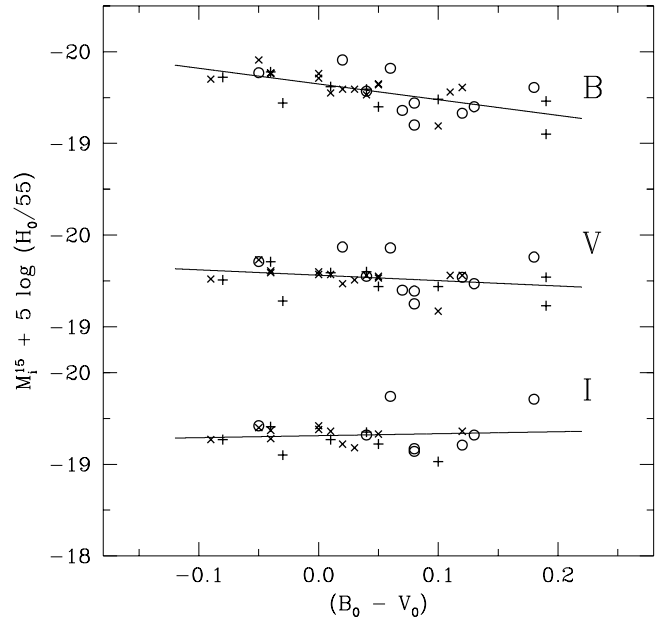,height=10.0cm}}}
\caption{Correlation of $(B_0 - V_0)$ color with the decay-rate
corrected magnitudes $M_i^{15}$.  The lines are equations (20) and (21)
of the text.  The shallow slope coefficients in $B$ and $V$, and the
reverse correlation of color with luminosity for the $I$ band data,
show that the correlations are not due to absorption and reddening,
but are an intrinsic property of the supernovae. \label{fig13}}
\end{figure*}

Figure~\ref{fig13} shows that there is still a dependence of the
corrected $M_i^{15}$ magnitudes on color that is not removed by the
$\Delta m_{15}$ corrections, contrary to the removal of the dependence
on Hubble type. The symbols are the same as in Fig.~\ref{fig12}.

The ordinates are the kinematic ($H_0 = 55$) absolute magnitudes,
$M_i^{15}$, based on variable local values of $H_0$ and the decay rate
corrections from the last subsection. The abscissa is the observed
$(B_0 - V_0)$ color that can be derived from columns (4) and (5) of
Table~\ref{tbl6}.

The equations for the least squares solutions for the lines are

\begin{equation}
M_B^{15} = 1.712\cdot (B_0 - V_0) - 19.648\,, \\
\end{equation}
\begin{equation}
M_V^{15} = 0.587\cdot (B_0 - V_0) - 19.562\,,
\end{equation}
and
\begin{equation}
M_I^{15} = -0.216\cdot (B_0 - V_0) - 19.312\,.
\end{equation}

\begin{figure*}
\centerline{\hbox{\psfig{figure=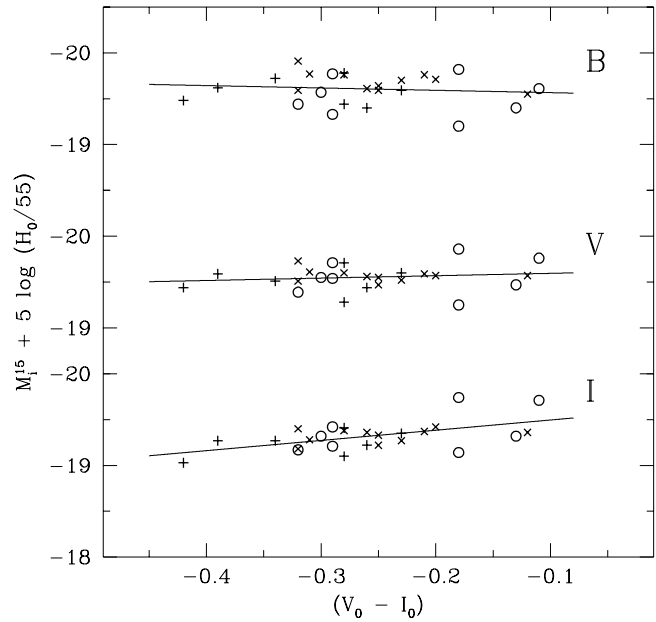,height=10.0cm}}}
\caption{Same as Fig.~\ref{fig13} but for $(V_0 - I_0)$
colors.  Note the reverse correlation in $V$ and $I$, where the
absolute magnitude is brighter for redder colors, showing again that
the correlation is not due to canonical extinction and reddening.
\label{fig14}}
\end{figure*}

Figure~\ref{fig14} is the same as Fig.~\ref{fig13} but with $(V_0-
I_0)$ colors derived from columns (10) and (11) of
Table~\ref{tbl6}. The least squares solutions for the lines are

\begin{equation}
M_B^{15} = 0.257\cdot (V_0 - I_0) - 19.540\,, \\
\end{equation}
\begin{equation}
M_V^{15} = -0.260\cdot (V_0 - I_0) - 19.621\,,
\end{equation}
and
\begin{equation}
M_I^{15} = -1.115\cdot (V_0 - I_0) - 19.607\,.
\end{equation}

The immediate consequence of equations (20) to (25) is that these
color dependencies {\it are not due to reddening and absorption}
because the slope coefficients in $B$ and $V$, nor their ratios,
conform to the canonical normal interstellar values of $A_B/E(B-V) =
4$, $A_V/E(B-V) = 3$, and $A_I/E(B-V) = 1.76$, and $A_B/E(V-I)= 3.2$,
$A_V/E(V-I) = 2.4$, and $A_I/E(V-I) = 1.4$ (\cite{schef82} 1982).  If,
for example, the variation of absolute magnitude with color index from
SN to SN were due entirely to absorption and reddening, then the slope
coefficients in the correlations of Fig.~\ref{fig13}, (Eqs.\,20 to 22)
would have to be $dM_B/E(B-V) = 4$, $dM_V/E(B-V) = 3$, and
$dM_I/E(B-V) = 1.76$, based on $A_V/A_I = 1.7$.  Not only are the
coefficients for $B$ and $V$ in equations (20) and (21) very different
from these requirements for reddening, but even the sign of the
variation in $I$ with $(B-V)$ is negative. This is impossible if the
cause is internal reddening with its corresponding dimming.

The situation is even more decisive from equations (23) to (25) using
$(V-I)$ colors in Fig.~\ref{fig14}. From E(V-I)/E(B-V) = 1.25 (Eq.\,1
of \cite{dea78} 1978), the predictions are $dM_B/E(V-I) = 3.2$,
$dM_V/E(V-I) = 2.4$, and $dM_I/E(V-I) = 1.41$.  These are not only far
different from the coefficients in equations (23) to (25), but again
the sense of the observed correlation in equations (24) and (25)
belies a reddening/extinction explanation. {\it Brighter} magnitudes
for redder colors are not possible for any known reddening and
absorption law.

The conclusion is that the correlations in Figs.~\ref{fig13} and
\ref{fig14} are due to intrinsic properties internal to the physics of
the SNe themselves rather than to reddening and absorption.  Hence,
any corrections to the absolute magnitudes based solely on assumed
absorption-to-reddening ratios for normal extinction are most likely
to be incorrect.

Nevertheless, the correlations in Figs.~\ref{fig13} and \ref{fig14}
are definite, and can be used to further reduce the absolute
magnitudes to some fiducial value of observed color.  We thus reduce
to a fiducial color of $(B_0-V_0) = 0.00$ as a second parameter, in
addition to the decay-rate parameter, in agreement with \cite{tri98}
(1998).

Our approach is different in principle from that of \cite{phi99}
(1999), where they assert that the color range at given decline rate
is due to reddening, and proceed accordingly. Our conclusion from the
above analysis is that for our sample of SNe~Ia, with $(B_{\rm max} -
V_{\rm max}) < 0.20$, the color spread at given decline rate is
intrinsic in nature, since correlation of the peak brightness with the
residual color for this sample is different from what is expected from
reddening.

Applying the color-term corrections of equations (20) and (21) to the
$M_B$ and $M_V$ magnitudes of Table~\ref{tbl6}, and taking the mean
values over the complete sample of 34 SNe~Ia in Table~\ref{tbl6} where
the decay rates are known, gives
\begin{equation}
   \langle{M_B^{\rm corr}}\rangle = \langle{M_B^{15} + ~{\rm
   color~term}}\rangle = -19.64 \pm 0.026\,,
\end{equation}
and
\begin{equation}
   \langle{M_V^{\rm corr}}\rangle = \langle{M_V^{15} + ~{\rm
   color~term}}\rangle = -19.61 \pm 0.026\,.
\end{equation}
These magnitudes are calculated with variable $H_0$ following the
precepts of \S 6.1 and reduced to $\Delta m_{15} = 1.1$ and $(B_0 -
V_0)= 0.00$.

For the calibrators in Table~\ref{tbl5} one finds in analogy
\begin{equation}
\langle{M_B^{\rm corr}}\rangle = -19.45 \pm 0.060\,,
\end{equation}
and
\begin{equation}
\langle{M_V^{\rm corr}}\rangle = -19.44 \pm 0.079\,.
\end{equation}
Comparison of equations (28) and (29) with equations (26) and (27)
shows that the $H_0 = 55$ assumption must be changed to accommodate
the differences of $0\fm19 \pm 0\fm08$ in $B$ and $0\fm17 \pm 0\fm09$
in $V$.  They require the global value of $H_0$, now fully corrected
for the three effects of (1) the change of $H_0$ outward, (2) the
light curve decay rate, and (3) the color variation with $M({\rm
max})$, to be
\begin{equation}
H_{0}(B) = 60.2 \pm 2\ksm\,,
\end{equation}
and
\begin{equation}
H_{0}(V) = 59.6 \pm 2\ksm\,.
\end{equation}
These values are only 2\% higher than equations (18) and (19) which
use only the decay-rate correction, and 1\% to 4\% higher than
equations (12) and (13) which are weighted towards SNe~Ia in spirals.

Other authors have adopted $\langle{M({\rm max})}\rangle$ values that
differ from those in equations (10) and (11) for the calibrator
absolute magnitudes.  For example, \cite{ken98} (1998), and
\cite{fre99} (1999) have discarded several of our calibrators, and
added two that are not based on direct Cepheid distances to the host
galaxy. They have consequently derived a fainter mean absolute $B$
magnitude than in equation (10).  Specifically, they assume that the
distances of the early-type galaxies NGC~1316 and NGC~1380 in the
Fornax cluster, parent galaxies to SN~1980N and SN~1992A, are
identical with that of the spiral NGC~1365 for which there is a
Cepheid distance. \cite{sun98} (1998) have also considered the
questionable SN~1980 and SN~1992A as possible calibrators.

However, there are reasons to suspect that NGC~1365 is in the
foreground of the Fornax cluster (\cite{satasa99} 1999), and therefore
that the precept of the fainter calibration used by \cite{ken98}
(1998), calibrating the two Fornax SNe~Ia via NGC~1365, is not
correct.  The evidence is that \cite{wel94} (1994) have demonstrated
that the multi-color light curves of SN~1989B in NGC~3627 and SN~1980N
in NGC~1316 are virtually identical, and in fact establish the
reddening and extinction to SN~1989B by comparing the magnitude shifts
in different passbands relative to SN~1980N.  Asserting then that
SN~1980N has the same peak brightness as SN~1989B yields the distance
modulus difference of $1\fm62 \pm 0\fm03$ (\cite{wel94} 1994).  There
is additional uncertainty of $\pm 0.17$ to allow for scatter in the
difference in peak brightness of two SNe~Ia with the same decline
rate. With our modulus of $(m - M)_0 = 30.22 \pm 0.12 $ (Eq.~7) for
SN~1989B, the derived modulus of NGC~1316, host to SN~1980N, is $31.84
\pm 0.21$.  This is $0\fm5$ more distant than the Cepheid distance of
NGC~1365 (\cite{mad98} 1998), but is close to the value found for the
early-type galaxies of the Fornax cluster by independent methods
(\cite{tam98b} 1998b).  In any case the implication of \cite{ken98}
(1998) and \cite{fre99} (1999) that the two SNe~Ia 1989B and 1980N
differ by $0\fm5$ in luminosity, which is based on the unproved
assertion that NGC~1365 and NGC~1316 are at the same distance, is not
credible.

The further consequence of the precept of equating the Cepheid
distance of NGC~1365 to that of the unknown distance to NGC~1316 and
NGC~1380 is that the decay-rate absolute magnitude relation which
\cite{fre99} (1999) deduce using their faint absolute magnitudes of
the two Fornax SNe~Ia is steeper than that in Fig.~\ref{fig12} by an
amount that compromises their conclusions concerning $H_0$, explaining
their abnormally high value of $H_0 \approx 73$.

\section{Summary and Conclusions}

     (1) Comparison of 12 epoch $HST$ frames in the $F555W$ band and
five epoch frames in the $F814W$ band has isolated 68 definite Cepheid
variables with periods between 3 and $\sim75$ days in the dust-rich
galaxy NGC~3627 in the Leo Group.

     (2) The adopted true modulus of $(m - M)_0 = 30.22 \pm 0.12$ that
results from analysis of various subsamples of the data according to
the degree of internal absorption, agrees well with the distance
moduli of $(m - M)_0 = 30.37 \pm 0.16$ for NGC~3368 (parent of
SN~1998bu) (\cite{tan95} 1995) and $(m - M)_0 = 30.01 \pm 0.19$ for
NGC~3351 (\cite{gra97} 1997), showing that the extended Leo Group in
fact exists (\cite{hum56} 1956, Table XI).

     (3) Combining the NGC~3627 Cepheid modulus of $(m -M)_0 = 30.22$
with the photometry of \cite{wel94} (1994) for the daughter SN~Ia
SN~1989B (corrected for extinction) gives the absolute magnitudes in
$B$ and $V$ at maximum light of $M_B = -19.36 \pm 0.18$, and $M_V =
-19.34 \pm 0.16$ for SN~1989B as listed in Table~\ref{tbl5}.

     (4) Combining these absolute magnitudes with those for the
previous six calibrators determined in previous papers of this series,
and with two additional calibrators recently available (SN~1974G in
NGC~4414 and SN~1998bu in NGC~3369), gives a mean calibration (without
second parameter corrections) of $\langle{M_B}\rangle = - 19.49 \pm
0.07$ and $\langle{M_V}\rangle = - 19.48 \pm 0.07$.

     (5) The absolute magnitudes $M_{\rm max}$ of blue SNe~Ia with
$(B_0 - V_0) <0.20$ correlate with the light curve decay rate $\Delta
m_{15}$. The correlation is approximated by a cubic equation which has
a flat plateau at intermediate values of $\Delta m_{15}$.  In
addition, $M_{\rm max}$ depends on the color $(B_0 - V_0)$. For the
derivation of the dependence on these second parameters, a slight
decrease of $H_0$ out to $10,000 \kms$ has been taken into
account. Once corrections for these two second parameters are applied,
no residual dependence of $M_{\rm max}$ on Hubble type is seen.

     (6) The calibrators have somewhat smaller mean values of $\Delta
m_{15}$ and bluer mean colors $(B_0 - V_0)$ than the distant SNe~Ia
defining the Hubble diagram. The application of second parameter
corrections therefore tends to increase $H_0$, but the effect is less
than 10\%.

     (7) The overall value of the Hubble constant determined here from
the current eight calibrators as applied to the distant SNe~Ia with
$1000 \kms < v < 30,000 \kms$ is $H_0 = 60 \pm 2$ in both $B$ and $V$
(Eqs. 30 and 31).

     (8) The Hubble constant derived here rests on Cepheid distances
whose zeropoint is set to an LMC modulus of 18.50. If the latter is
revised upwards by $\approx 0\fm06$ (\cite{fed98} 1998), the
consequent value of the Hubble constant is

\begin{equation}
H_0 = 58 \pm 2\ksm\,.
\end{equation}

\acknowledgments{Acknowledgment}

We thank the many individuals at STScI who worked hard behind the
scenes to make these observations possible, and wish to particularly
mention Doug van Orsow, George Chapman, Bill Workman, Merle Reinhardt
and Wayne Kinzel.  A.S. and A.S. acknowledge support from NASA through
grant GO-5427.02-93A from the Space Telescope Science Institute, which
is operated by the Association of Universities for Research in
Astronomy.  L.L. and G.A.T. thank the Swiss National Science
Foundation for continued support.

%
\begin{deluxetable}{lccc}
\tablecaption{Journal of Observations. \label{tbl1}}
\tablewidth{0pt} 
\tablehead{
\colhead{Data Archive Designation} &\colhead{HJD at Midexposure} &
\colhead{Filter} & \colhead{Exposure time (s)}}
\startdata
U35I0101R + ...02R & 2450765.07048 & $F555W$ & 4900 \nl
U35I0103R + ...04R & 2450765.20105 & $F814W$ & 5000 \nl
U35I1201R + ...02R & 2450768.96875 & $F555W$ & 4900 \nl
U35I0201R + ...02R & 2450773.53863 & $F555W$ & 4900 \nl
U35I0301R + ...02R & 2450780.59626 & $F555W$ & 4900 \nl
U35I0401R + ...02R & 2450786.37739 & $F555W$ & 4900 \nl
U35I0403R + ...04R & 2450786.51004 & $F555W$ & 4900 \nl
U35I0501R + ...02R & 2450791.28343 & $F814W$ & 5000 \nl
U35I0601R + ...02R & 2450796.66105 & $F555W$ & 4900 \nl
U35I0701R + ...02R & 2450801.23025 & $F555W$ & 4900 \nl
U35I0703R + ...04R & 2450801.36499 & $F814W$ & 5000 \nl
U35I0801R + ...02R & 2450804.39306 & $F555W$ & 4900 \nl
U35I0901R + ...02R & 2450810.10888 & $F555W$ & 4900 \nl
U35I0903R + ...04R & 2450810.24084 & $F814W$ & 5000 \nl
U35I01001R + ...02R & 2450817.03382 & $F555W$ & 4900 \nl
U35I01101R + ...02R & 2450823.88997 & $F555W$ & 4900 \nl
U35I01103R + ...04R & 2450824.02262 & $F814W$ & 5000 \nl
\enddata
\end{deluxetable}
\clearpage

\begin{deluxetable}{ccccrr}
\tablecaption{Mean Absolute Magnitudes Corrected for Decay Rate as
Function of Morphological Type based on $H_0~({\rm global}) =
55$\tablenotemark{a} \label{tbl7}}
\tablewidth{0pt}
\tablehead{
 \colhead{T} & \colhead{$\langle{M_B^{15}}\rangle$} & 
 \colhead{$\langle{M_V^{15}}\rangle$} & 
 \colhead{$\langle{M_I^{15}}\rangle$} & \colhead{$n(B, V)$} & 
 \colhead{$n(I)$} \\
 \colhead{(1)} & \colhead{(2)} & \colhead{(3)} & \colhead{(4)} & 
 \colhead{(5)} & \colhead{(6)}} 
\startdata
3, 4, 5   & $-19.54\pm0.078$ & $-19.58\pm0.071$ & $-19.38\pm0.088$ & 10 &  8 \nl
1, 2      & $-19.51\pm0.071$ & $-19.48\pm0.054$ & $-19.24\pm0.054$ &  9 &  7 \nl
$-$3 to 0 & $-19.64\pm0.043$ & $-19.54\pm0.032$ & $-19.32\pm0.024$ & 15 & 11 \nl
All       & $-19.57\pm0.034$ & $-19.54\pm0.028$ & $-19.32\pm0.032$ & 34 & 26 \nl
\multicolumn{6}{c}{Calibrators from Table 5} \nl
3, 4, 5   & $-19.43\pm0.047$ & $-19.39\pm0.071$ & \nodata          &  8 
& \nodata \nl
\enddata
\tablenotetext{a}{Listed are the magnitudes corrected for decay rate via 
equations (15), (16), and (17).} 
\end{deluxetable}
\clearpage

\clearpage

\clearpage

\clearpage
 
\end{document}